%% file: main.tex
\documentclass{article} %
\usepackage{iclr2027_conference,times}
\usepackage[utf8]{inputenc} %
\usepackage[T1]{fontenc}    %
\usepackage{hyperref}       %
\usepackage{xurl}           %
\usepackage{booktabs}       %
\usepackage{amsfonts}       %
\usepackage{nicefrac}       %
\usepackage{microtype}      %
\usepackage[svgnames,x11names,table]{xcolor}         %

\usepackage[textsize=tiny]{todonotes}

\input{math_commands}

\usepackage{graphicx}       %
\usepackage{subcaption}     %
\usepackage{multirow}
\usepackage{makecell}       %
\usepackage{colortbl}
\usepackage{bbm}
\usepackage{tikz}
\usepackage{pifont}

\usepackage{hhline}
\usepackage{framed}
\usepackage{float}
\usepackage{enumitem}
\usepackage{fvextra}        %
\fvset{fontsize=\scriptsize,breaklines=true,breakanywhere=true,xleftmargin=6pt}
\usepackage[capitalize,noabbrev]{cleveref}

\usepackage{mathtools}
\usepackage{dcolumn}
\newcolumntype{d}{D{.}{.}{-1}} %

\Crefname{section}{Section}{Sections}
\Crefname{table}{Table}{Tables}
\Crefname{figure}{Figure}{Figures}
\crefname{appendix}{Appendix}{Appendices}
\Crefname{appendix}{Appendix}{Appendices}
\crefname{subappendix}{Appendix}{Appendices}
\Crefname{subappendix}{Appendix}{Appendices}

\hypersetup{
    colorlinks = true,
    linkcolor = red,
    anchorcolor = red,
    urlcolor = red,
    citecolor = MediumBlue,
}

\usepackage{xcolor}
\usepackage{xspace}
\makeatletter
\DeclareRobustCommand\onedot{\futurelet\@let@token\@onedot}
\def\@onedot{\ifx\@let@token.\else.\null\fi\xspace}

\makeatother
\usepackage{wrapfig}

\newcommand{\moesq}{\textbf{\textsc{m}}o\textbf{\textsc{esq}}\xspace}
\newcommand{\ours}{\moesq}

\usepackage{xparse}
\NewDocumentCommand{\alignednum}{m o}{%
  \makebox[5em][c]{%
    \makebox[0pt][r]{#1}%
    \makebox[0pt][l]{%
        \IfValueTF{#2}
          {$_{\pm#2}$}
          {\phantom{$_{\pm0.00}$}}
    }%
  }%
}

\title{Hardware-Native Joint Sparse-Quantization for Trillion-Scale Mixture-of-Experts}

\newif\ifpreprint
\newcommand{\preprint}{\preprinttrue\iclrfinalcopy}
\AtBeginDocument{\ifpreprint
  \let\iclrmaketitle\maketitle
  \renewcommand{\maketitle}{\iclrmaketitle\lhead{Preprint}}%
\fi}

\preprint      %

\author{Kwanhee Lee\thanks{Work done during an internship at ISTA.} \\
POSTECH \& ISTA \\
\And
Namhoon Lee \\
POSTECH \\
\And
Dan Alistarh\thanks{Correspondence to: \texttt{dan.alistarh@ist.ac.at}.} \\
ISTA \\
}

\begin{document}

\maketitle

\input{texts/0_abstract}
\input{texts/1_intro}

\input{texts/2_background}

\input{texts/3_method}
\input{texts/4_experiments}
\input{texts/5_conclusion}
\ificlrfinal\input{texts/ack}\fi %
\bibliography{reference}
\bibliographystyle{iclr2027_conference}
\input{texts/6_appendix}

\end{document}

%% file: math_commands.tex
\usepackage{amsmath, amsthm, amsopn, amstext, amsfonts, bm, algorithm, algpseudocode, graphicx, mathabx}

\def\eqref#1{equation~\ref{#1}}

\def\1{\bm{1}}

\DeclareMathAlphabet{\mathsfit}{\encodingdefault}{\sfdefault}{m}{sl}
\SetMathAlphabet{\mathsfit}{bold}{\encodingdefault}{\sfdefault}{bx}{n}

%% file: texts/0_abstract.tex
\begin{abstract}

Mixture-of-Experts (MoE) architectures allow frontier language models to scale toward trillions of parameters, but their deployment is constrained by massive memory footprints and bandwidth limits. 
While modern accelerators feature Sparse Tensor Cores (SpTCs) to reduce weight storage and boost throughput via low-precision semi-structured sparsity, exploiting them in MoEs is hindered by severe degradation of model quality and a lack of grouped sparse GEMM primitives.
We present \ours, an end-to-end hardware–software co-design framework that compresses expert weights into hardware-native low-precision sparse representations and accelerates their execution on SpTC. 
Algorithmically, \ours relaxes discrete semi-structured support selection via continuous reparameterization, enabling differentiable joint optimization with quantized weights under a router-weighted reconstruction objective scaled via expert-parallel compression. 
Systemically, we implement a custom grouped sparse GEMM kernel tailored for sparse, low-precision grouped GEMM on SpTC for MoE inference. 
Across MoE scales spanning 30B to frontier-scale trillion parameters, \ours advances state-of-the-art joint sparse-quantization by up to $4.35$ percentage points in task accuracy, preserving $96.09\%$ of original model's performance.
System-level benchmarks on NVIDIA B200 GPUs show that our kernel outpaces vendor baseline by up to $1.65\times$, driving $1.18\times$ higher serving throughput and up to a $4.03\times$ drop in end-to-end latency.
Ultimately, these results establish hardware--software co-design as a practical and necessary pathway for scalable, highly efficient MoE deployment.

\end{abstract}

%% file: texts/1_intro.tex
\section{Introduction}
\label{sec:intro}

Recent advances in Mixture-of-Experts (MoE) architectures have emerged as a dominant paradigm for scaling foundation models, pushing parameter counts toward trillions while keeping per-token inference FLOPs manageable through dynamic expert routing~\citep{kimiteam2026k25,deepseekai2026v4,qwen2026qwen38}.
However, this architectural efficiency introduces severe deployment bottlenecks in real-world serving.
Despite activating only a fraction of weights per token, serving massive MoEs typically requires hosting the entire pool of experts in high-bandwidth device memory, leading to prohibitive memory footprints and memory bandwidth limitations under varying batch regimes.
Consequently, the sheer storage and memory-bound computational demands of these billions-to-trillions of expert parameters severely limit the practical throughput and cost efficiency of MoE deployment.

A promising yet underutilized opportunity to overcome these memory and compute constraints lies in modern hardware accelerators.
Modern architectures feature Sparse Tensor Cores (SpTC) that natively support fine-grained semi-structured ($N:M$) sparse low-precision matrix multiplications, offering a principled path to simultaneously compress weight storage and accelerate computational throughput~\citep{nvidia2021a100datasheet,nvidia2022h100datasheet,nvidia2024dgxb200datasheet}.
Despite this capability, the prevailing body of research and deployment efforts for large-scale MoEs has disproportionately concentrated on dense, low-precision quantization and execution~\citep{huang2025mcmoe,fu2025eaquant,dadgarnia2026gsq,egiazarian2026bridging,nvidia2026qwen35nvfp4,nvidia2026kimik25nvfp4}.
As a result, the synergistic potential of hardware-accelerated semi-structured sparsity remains largely unexplored in the context of massive MoE serving.

Realizing these benefits in large-scale MoEs, however, poses two fundamental challenges.
First, preserving model fidelity under the compound constraints of semi-structured sparsity and low-precision quantization is remarkably difficult.
Prior work has explored joint sparsification and quantization~\citep{frantar2023sparsegpt,guo2024jsq,guo2026obr}, but extending these techniques to trillion-parameter MoEs remains largely unproven, as the rigid semi-structured pattern combined with aggressive quantization exacerbates degradation across diverse expert representations.
Second, even when an accurate sparse-quantized model is obtained, translating theoretical reductions into wall-clock speedups is fundamentally bottlenecked by the software execution stack.
MoE inference relies on grouped GEMM to efficiently batch variable routing-dependent tokens across experts; however, existing grouped GEMM kernels exclusively target dense formats, leaving hardware-accelerated semi-structured sparse low-precision operations unsupported~\citep{cutlass_blackwell_docs}.

In this work, we propose \ours, an end-to-end framework that bridges algorithm design and hardware acceleration for sparse, low-precision MoE serving.
At the algorithmic level, \ours overcomes the combinatorial complexity of rigid semi-structured support selection by relaxing discrete mask decisions through continuous reparameterization.
This enables differentiable, joint adaptation of sparse supports and weights under a quantization-aware reconstruction objective.
To scale seamlessly to trillion-parameter architectures, \ours weights expert reconstruction by dynamic routing significance and orchestrates the compression pipeline across distributed experts in parallel.
Crucially, to turn these compact representations into tangible deployment gains, we develop a specialized grouped sparse GEMM kernel that executes native semi-structured low-precision operations on SpTC.
By co-designing the compression objective with dedicated kernel execution, \ours achieves both high model fidelity and substantial end-to-end efficiency gains on modern hardware.

Extensive evaluations across MoE architectures spanning from 30B to frontier trillion-parameter scales demonstrate that \ours sets a new standard in compression fidelity.
Specifically, \ours improves average task accuracy by up to $4.35$ percentage points over the strongest joint sparsification and quantization baselines, retaining $96.09\%$ of the uncompressed model’s accuracy even at the trillion-parameter scale.
At the systems level, our grouped sparse GEMM kernel for NVIDIA B200 GPUs achieves up to a $1.65\times$ speedup over its state-of-the-art dense counterpart.
Under matched concurrency, this kernel acceleration translates into a $1.18\times$ higher end-to-end serving throughput over the fastest dense low-precision backends, up to a $4.03\times$ reduction in end-to-end latency compared to the INT4 Marlin backend~\citep{frantar2025marlin}, and the lowest time-per-output-token across all evaluated systems.
Taken together, these results demonstrate that our hardware--software co-design enables high-fidelity compression while reducing memory footprint and improving serving efficiency for large-scale MoE inference.

%% file: texts/2_background.tex
\section{Background}
\label{sec:background}

A sparsely activated Mixture-of-Experts (MoE) layer replaces the dense feed-forward network with $E$ experts and routes each token to only $k\ll E$ of them~\citep{shazeer2017moe,lepikhin2021gshard,fedus2022switch}, decoupling total parameter count from per-token compute~\citep{krajewski2024moescaling,abnar2025sparsity,nakamura2026optimalsparsity}.
Recent open-weight MoEs span roughly 750B--2.8T total parameters while activating only 32B--104B parameters per token~\citep{zai2026glm52,lgai2026kexaone2,kimiteam2026k25,deepseekai2026v4,qwen2026qwen38,moonshot2026k3}.
Although sparse routing limits per-token computation, deployment must store and make available the full expert pool, causing expert-weight memory to scale with total rather than active parameters.
Kimi-K2.5 illustrates this disparity: the model contains 1T total parameters but activates only 32B per token, while its released packed INT4 checkpoint occupies approximately 595~GB~\citep{kimiteam2026k25,vllm2026kimik25recipe}.

Modern hardware accelerators provide an opportunity to reduce both memory and computation by operating directly on sparse low-precision representations.
Sparse Tensor Cores (SpTCs) exploit semi-structured $N{:}M$ sparsity by encoding nonzero values with sparse metadata and skipping operations on structured zeros, while simultaneously supporting low-precision data formats~\citep{mishra2021nm,ocp2023mx,rouhani2023mx,nvidia2025nvfp4}.
Because this path reduces arithmetic, vendors commonly use sparse throughput to report the maximum performance of recent accelerators~\citep{nvidia2021a100datasheet,nvidia2022h100datasheet}.
For instance, on the NVIDIA Blackwell architecture, block-scaled FP4 formats (NVFP4 and MXFP4) can be coupled with \emph{paired $4{:}8$ sparsity} on fifth-generation SpTCs, providing up to 18~PFLOP/s for sparse NVFP4, twice the throughput of the dense NVFP4 path~\citep{cutlass_blackwell_docs,nvidia2024dgxb200datasheet}.
Realizing this opportunity in routed MoE inference requires both expert weights that satisfy the exact paired sparsity, microscaling, and metadata constraints of the hardware representation and an execution path that can operate on this representation across dynamically routed experts.

The first requirement can be addressed by combining sparsity and quantization, an approach studied primarily to push model compression beyond standalone sparsification~\citep{sun2024wanda,lee2026elsa} or quantization~\citep{frantar2023gptq,lin2024awq,tseng2024qtip,dadgarnia2026gsq}.
SparseGPT~\citep{frantar2023sparsegpt} sparsifies LLMs through second-order error compensation and can subsequently quantize the remaining weights for further compression.
The interaction between sparsification and quantization has also been analyzed~\citep{harma2025interplay,kim2026order}, and more recent methods explicitly account for it: JSQ~\citep{guo2024jsq} uses a quantization-aware pruning metric and activation editing to reduce quantization error, while OBR~\citep{guo2026obr} rotates, sparsifies, compensates, and then quantizes the weights.
Because these methods were developed and evaluated primarily for compressing dense LLMs, their solution quality under the target hardware-native constraints and their scalability to trillion-parameter MoE models remain unclear.

Producing hardware-compliant sparse-quantized weights addresses only the first requirement; these weights must also be executed efficiently under routed MoE workloads.
Modern GPU libraries provide single-problem GEMM kernels that combine semi-structured sparsity with low-precision arithmetic, including $2{:}4$ INT8 on Ampere, $2{:}4$ FP8 on Hopper, and paired $4{:}8$ NVFP4 on Blackwell~\citep{mishra2021nm,nvidia2022h100datasheet,cutlass_blackwell_docs}.
Routed MoE inference, however, relies on grouped GEMM to evaluate the expert products $Y_e=X_eW_e^\top$ in a single launch even though routing gives each expert a different token dimension $N_e=|X_e|$.
Although low-precision grouped GEMM kernels, including NVFP4 grouped GEMM, are available, no publicly available sparse grouped GEMM path directly supports the target paired $4{:}8$ NVFP4 representation~\citep{nvidia2025cutlassgroupednvfp4}.
Hardware-native compression alone is therefore insufficient for efficient MoE execution; realizing its benefits also requires a sparse low-precision grouped GEMM kernel that directly processes the compressed expert weights.

Taken together, these gaps leave the capabilities of SpTCs disconnected from modern MoE deployment: no established path simultaneously preserves model fidelity for trillion-parameter scale MoE models and translates sparse low-precision hardware support into reduced expert storage and accelerated routed execution.
Bridging this divide requires hardware--software co-design that aligns MoE compression and routed execution with hardware-native representation constraints.

%% file: texts/3_method.tex
\section{Method}
\label{sec:method}
\label{sec:method_joint}

We propose \ours, which consists of an algorithmic procedure that learns hardware-native sparse-quantized representations for expert weights and a grouped sparse GEMM that executes the same representation on SpTC.
We first describe how \ours jointly adapts sparse supports and quantized values for MoE models, then detail its implementation and corresponding execution kernel.

\subsection{Problem Formulation}
We seek a compressed parameterization $\widetilde{W}$ that minimizes the reconstruction error of a targeted module $f$ (e.g., a linear projection or an MoE expert) evaluated on calibration inputs $X$:
\begin{equation}
\min_{\widetilde{W}}\ \mathcal{L}(\widetilde{W})
:=
\big\|f(X; W) - f(X; \widetilde{W})\big\|_F^2
\quad\text{s.t.}\quad
\widetilde{W} \in \mathcal{C} := \mathcal{S} \cap \mathcal{Q},
\label{eq:feasible}
\end{equation}
where $\mathcal{S}$ denotes the set of hardware-admissible semi-structured sparsity patterns (e.g., paired 4:8), and $\mathcal{Q}$ denotes the set of admissible quantized values accompanied by their associated shared scaling factors (e.g., NVFP4).
Because support selection, quantized representation, and shared scale estimation are tightly coupled, \cref{eq:feasible} constitutes a challenging discrete combinatorial optimization problem.

\subsection{Learning Hardware-Native Sparse-Quantized Representations}
To establish a differentiable formulation, we relax each discrete support selection into a continuous distribution over admissible patterns via the Gumbel--Softmax reparameterization trick~\citep{jang2017gumbel, maddison2017concrete, fang2024maskllm}.

Consider a weight matrix $W \in \mathbb{R}^{N \times K}$ partitioned into $R$ structured groups of size $m$.
Each group $W_{[r]} \in \mathbb{R}^{m}$ is constrained to take a pattern from a predefined set of admissible binary vectors $\mathcal{P} = \{\rho_1, \dots, \rho_P\} \subset \{0,1\}^{m}$.
For example, under a paired 4:8 sparse format ($m=8, P=6$), each pattern $\rho_p$ retains two non-zero pairs of weights (see \cref{app:sq_format} for full specifications).
Let $A \in \mathbb{R}^{R \times P}$ represent a logit matrix scoring the candidate patterns, where $A_{r,p}$ measures the affinity of group $r$ for pattern $\rho_p$.
At optimization step $t$, we sample independent and identically distributed (i.i.d.) Gumbel noise to construct the relaxed support mask:
\begin{equation}
G_{t,r,p} \overset{\mathrm{iid}}{\sim} \operatorname{Gumbel}(0,1),
\qquad
\pi_{t,r,:} = \operatorname{softmax}\!\left(\frac{A_{r,:} + G_{t,r,:}}{\tau_t}\right),
\qquad
Z_{t,[r]} = \sum_{p=1}^{P} \pi_{t,r,p} \rho_p,
\label{eq:gumbel}
\end{equation}
where $\tau_t > 0$ denotes a temperature hyperparameter governing the softness of the categorical distribution $\pi_{t,r,:}$.
This continuous relaxation provides a differentiable surrogate for support selection, enabling gradient-based updates to the pattern logits $A$.
Annealing $\tau_t$ toward zero over training drives the relaxed support mask $Z_{t,[r]}$ toward a deterministic, one-hot selection of an admissible pattern.
Although the Gumbel--Softmax relaxation renders support selection differentiable, it does not implicitly enforce the quantization constraint $\mathcal{Q}$.
To address this, we introduce an auxiliary full-precision matrix $V \in \mathbb{R}^{N \times K}$ and couple it with the relaxed support mask to form the soft-masked weight matrix $Y_t := Z_t \odot V$.

A target quantization operator $\mathcal{Q}_w$ subsequently projects $Y_t$ into $\mathcal{Q}$.
Let $\mathcal{G}_w$ denote the admissible target value grid.
For each quantization block $Y_{t,[b]}$, the operator $\mathcal{Q}_w$ derives a local scaling factor $s_{t,b}$ according to the target precision format and projects normalized weights onto $\mathcal{G}_w$:
\begin{equation}
s_{t,b} := \operatorname{Scale}_w(Y_{t,[b]}),
\qquad
\mathcal{Q}_w(Y_t)_{[b]} := s_{t,b} \, \Pi_{\mathcal{G}_w}\!\left(\frac{Y_{t,[b]}}{s_{t,b}}\right),
\label{eq:fakequant}
\end{equation}
where $\operatorname{Scale}_w(\cdot)$ is the scale-computation function (e.g., maximum absolute value), and $\Pi_{\mathcal{G}_w}(\cdot)$ denotes the nearest-neighbor projection onto grid $\mathcal{G}_w$.

During backward passes, we treat scaling factors $s_{t,b}$ as constants and employ the Straight-Through Estimator (STE)~\citep{bengio2013ste} to propagate gradients through the non-differentiable projection $\Pi_{\mathcal{G}_w}$.
Consequently, the reconstruction loss jointly updates the support logits $A$ and auxiliary weights $V$, ensuring that candidate supports are evaluated directly under their quantization-constrained realization.

\subsection{Scalable Blockwise Expert Compression}
To scale continuous optimization across MoE models, we follow a blockwise per-expert decomposition inspired by GSQ~\citep{dadgarnia2026gsq}.
At block $l$, the router assigns set of token representations $X_{l,e}$ to expert $e$ and their corresponding gating coefficients into the aligned vector $w_{l,e}$.
Evaluating the uncompressed expert on these inputs provides the reconstruction target $Y_{l,e}=f_{l,e}(X_{l,e};\theta_{l,e})$.
We reconstruct each expert as a unified gate--up--down block $\theta_{l,e}=(W_{\mathrm{gate}},W_{\mathrm{up}},W_{\mathrm{down}})$, with separate support logits and latent weights for each projection and dynamic activation quantization $\mathcal{Q}_a$ inside the reconstruction loop.
Writing $f^{\mathcal{Q}_a}_{l,e}(X;\widetilde{\theta})$ for activation-quantized expert execution, we optimize the router-weighted reconstruction objective:
\begin{equation}
\mathcal{L}^{\mathrm{rw}}_{l,e}(\widetilde{\theta})
:=
\big\|\operatorname{diag}(w_{l,e})
\big(Y_{l,e}-f^{\mathcal{Q}_a}_{l,e}(X_{l,e};\widetilde{\theta})\big)\big\|_F^2.
\label{eq:diag}
\end{equation}
This objective scales each token residual by the same gate coefficient that weights the expert output in the routed MoE layer.
As summarized in \cref{alg:moesq}, experts within a block are optimized independently across devices, after which the finalized block generates calibration inputs for block $l+1$.

\begin{algorithm}[t]
\caption{Hardware-Native Expert Compression for Mixture-of-Experts}
\label{alg:moesq}
\begin{algorithmic}[1]
\Require MoE model with $L$ blocks and $E$ experts; calibration data $\mathcal{D}$; pattern set $\mathcal{P}$; quantizers $\mathcal{Q}_w, \mathcal{Q}_a$; temperature schedule $\{\tau_t\}_{t=1}^{T}$; optimization steps $T$
\State $H_1 \leftarrow \operatorname{Embed}(\mathcal{D})$
\For{$l = 1, \ldots, L$}
    \State $\{X_{l,e}, w_{l,e}\}_{e=1}^{E} \leftarrow \operatorname{Route}_l(H_l)$
    \ForAll{$e \in \{1, \ldots, E\}$ \textbf{in parallel}}
        \State $Y_{l,e} \leftarrow f_{l,e}(X_{l,e}; \theta_{l,e})$
        \State Initialize $\{A^{(j)}, V^{(j)}\}_{j \in \{\mathrm{gate}, \mathrm{up}, \mathrm{down}\}}$
        \For{$t = 1, \ldots, T$}
            \For{$j \in \{\mathrm{gate}, \mathrm{up}, \mathrm{down}\}$}
                \State Sample noise $G_t^{(j)}$ and construct $Z_t^{(j)}$ via \cref{eq:gumbel}
                \State $Y_t^{(j)} \leftarrow Z_t^{(j)} \odot V^{(j)}$
                \State $\widetilde{W}_t^{(j)} \leftarrow \mathcal{Q}_w(Y_t^{(j)})$ via \cref{eq:fakequant}
            \EndFor
            \State $\widetilde{\theta}_t \leftarrow \{\widetilde{W}_t^{(j)}\}_{j \in \{\mathrm{gate}, \mathrm{up}, \mathrm{down}\}}$
            \State Update parameters $\{A^{(j)}, V^{(j)}\}_j$ using $\nabla \mathcal{L}^{\mathrm{rw}}_{l,e}(\widetilde{\theta}_t)$
        \EndFor
        \State $Z_{[r]}^{\star(j)} \leftarrow \rho_{\operatorname{argmax}_p A_{r,p}^{(j)}}$ for every group $r$ and projection $j$
        \State $\theta^{\star}_{l,e} \leftarrow \{\mathcal{Q}_w(Z^{\star(j)} \odot V^{(j)})\}_{j \in \{\mathrm{gate}, \mathrm{up}, \mathrm{down}\}}$
    \EndFor
    \State $H_{l+1} \leftarrow \operatorname{Block}_l(H_l; \{\theta^{\star}_{l,e}\}_{e=1}^{E})$
\EndFor
\State \Return $\{\theta^{\star}_{l,e}\}_{l,e}$
\end{algorithmic}
\end{algorithm}

\subsection{Implementation Details}
We keep the router and all non-expert parameters fixed during compression.
For each block $l$, we route the current calibration states $H_l$ once and cache the expert inputs $X_{l,e}$, routing weights $w_{l,e}$, and uncompressed targets $Y_{l,e}$ for reuse throughout refinement.
The experts in a block then form independent optimization problems: we shard them across devices, optimize the assigned experts locally, and synchronize only after all experts have been finalized.
The resulting sparse-quantized experts are assembled into the block and applied to $H_l$ to produce the calibration states for block $l+1$, so each block observes the error accumulated by the compressed prefix.

We initialize the support logits and latent weights from an SGPTQ solution~\citep{frantar2023sparsegpt,frantar2023gptq} and recompute quantization scales from the current soft-masked weights on every refinement forward pass.
For NVFP4~\citep{nvidia2025nvfp4}, the block and tensor scales are computed as
\begin{equation}
\operatorname{Scale}_w(Y_{t,[b]}) = \frac{\Pi_{\mathrm{E4M3}}\!\left(\gamma_t \|Y_{t,[b]}\|_\infty / 6\right)}{\gamma_t},
\qquad
\gamma_t = \frac{448}{\max_b \|Y_{t,[b]}\|_\infty / 6},
\label{eq:nvfp4-scale}
\end{equation}
where $6$ and $448$ are the maximum magnitudes of the FP4-E2M1 value grid and FP8-E4M3 scale format, respectively.
We stop gradients through these scales, while the dynamic activation quantizer independently computes scales for the block input and the intermediate SwiGLU activation.
After refinement, we select the highest-logit pattern in every sparsity group, recompute the final scales, and pack the weights as E2M1 values, sparse metadata, and one UE4M3 scale per $32$ logical weights.
Full optimization settings are reported in \cref{app:setup}.

\paragraph{Grouped sparse GEMM execution on SpTC.}
\label{sec:kernel}
A routed MoE layer evaluates a collection of independent expert products $Y_e=X_eW_e^\top$, whose token dimensions $N_e$ vary with the routing decisions.
Grouped GEMM generalizes batched GEMM by allowing these problems to have different matrix dimensions, transpositions, and scaling factors while executing them within one kernel launch~\citep{hejazi2024groupedgemm}.
This organization avoids a separate launch for each expert while preserving its routing-dependent problem shape.

To extend grouped execution to support paired $4{:}8$ NVFP4 weights, we combine the Blackwell block-scaled sparse MMA mainloop with the grouped-problem abstraction and persistent scheduler that CUTLASS exposes separately~\citep{cutlass_blackwell_docs,nvidia2025cutlassgroupednvfp4}.
Because the sparse MMA requires the sparse matrix as its left operand $A$, we place $W_e\in\mathbb{R}^{M\times K}$ on $A$ and compute $Y_e^\top=W_eX_e^\top$~\citep{cutlass_blackwell_docs}.
This transformation is zero-copy: row-major $X_e\in\mathbb{R}^{N_e\times K}$ has the same memory layout as the column-major operand $B$, and the column-major $M\times N_e$ result has the row-major layout expected for $Y_e$.

We pack the exported weights offline as E2M1 values, sparse metadata, and one UE4M3 scale per $32$ logical weights, and quantize expert activations at run time into corresponding $32$-element scale blocks using $\mathcal{Q}_a$.
The sparse MMA applies the block scales of both operands, while the epilogue combines their tensor-wide scales into a per-expert factor $\alpha_e$ before casting the FP32 accumulator to BF16.
Further details on kernel configuration, activation quantization, and serving integration are provided in \cref{app:kernel_impl}.

%% file: texts/4_experiments.tex
\section{Experiments}
\label{sec:experiments}

We evaluate whether \ours addresses both requirements for hardware-native sparse MoE deployment: preserving model fidelity under the target representation and translating this representation into kernel and end-to-end serving gains.
We then additionally compare its accuracy--memory--serving tradeoff with other state-of-the-art expert-compression methods.

\begin{table}[t]
\centering
\caption{OpenLLM Leaderboard v1 results across three MoE scales under the paired $4{:}8$ NVFP4 W4A4 expert representation.
  Recovery (Rec.) is relative to each model's original average accuracy; bold marks the best compressed result.}
\label{tab:main}

\small
\setlength{\tabcolsep}{4pt}
\resizebox{\linewidth}{!}{%
\begin{tabular}{llcccccccc}
\toprule
Model & Method & ARC-c & GSM8K & MMLU & WG & HSwag & TQA & Avg & Rec. \\
\midrule
\multirow{6}{*}{Kimi-K2.5}

  & Reference     & 74.06 & 94.39 & 89.53 & 81.93 & 91.96 & 62.69 & 82.43 & 100\% \\
  & Dense NVFP4   & 73.21 & 93.35 & 89.23 & 81.45 & 91.87 & 62.29 & 81.90 & 99.36\% \\
\cmidrule(lr){2-10}
  & JSQ           & 59.22 & 68.92 & 71.87 & 73.88 & 80.95 & 50.58 & 67.57 & 81.97\% \\
  & SGPTQ         & 63.05 & 79.30 & 82.10 & 79.72 & 84.39 & 54.92 & 73.91 & 89.67\% \\
  & OBR           & 63.65 & 82.79 & 82.49 & 80.35 & 85.34 & 55.78 & 75.07 & 91.07\% \\
  & Ours          & \textbf{69.97} & \textbf{88.70} & \textbf{85.34} & \textbf{82.40} & \textbf{88.57} & \textbf{60.28} & \textbf{79.21} & \textbf{96.09\%} \\
\midrule
\multirow{6}{*}{Qwen3.5-397B-A17B}
  & Reference     & 79.10 & 85.06 & 89.49 & 76.16 & 90.85 & 60.97 & 80.27 & 100\% \\
  & Dense NVFP4   & 79.15 & 78.85 & 89.49 & 75.95 & 90.75 & 59.59 & 78.96 & 98.37\% \\
\cmidrule(lr){2-10}
  & JSQ           & \textbf{75.51} & 4.62 & 84.29 & 78.06 & 82.43 & 52.80 & 62.95 & 78.43\% \\
  & SGPTQ         & 74.74 & 49.81 & 85.67 & 78.77 & 81.02 & 55.15 & 70.86 & 88.27\% \\
  & OBR           & 75.00 & 65.28 & 86.00 & \textbf{78.85} & 81.89 & \textbf{55.30} & 73.72 & 91.84\% \\
  & Ours          & 74.49 & \textbf{71.65} & \textbf{86.97} & 78.30 & \textbf{84.74} & 55.10 & \textbf{75.21} & \textbf{93.70\%} \\
\midrule
\multirow{6}{*}{Qwen3-30B-A3B}
  & Reference     & 69.45 & 89.54 & 79.34 & 69.61 & 77.92 & 53.78 & 73.27 & 100\% \\
  & Dense NVFP4   & 67.80 & 88.32 & 78.04 & 70.38 & 77.15 & 52.52 & 72.37 & 98.77\% \\
\cmidrule(lr){2-10}
  & JSQ           & 52.13 & 65.88 & 60.70 & 62.19 & 51.68 & 43.14 & 55.96 & 76.37\% \\
  & SGPTQ         & 56.14 & 72.86 & 66.51 & 65.98 & 56.33 & 48.72 & 61.09 & 83.37\% \\
  & OBR           & 56.23 & 72.40 & 66.60 & 65.04 & 57.50 & 48.76 & 61.09 & 83.37\% \\
  & Ours          & \textbf{59.73} & \textbf{79.45} & \textbf{69.89} & \textbf{67.80} & \textbf{63.60} & \textbf{52.15} & \textbf{65.44} & \textbf{89.30\%} \\
\bottomrule
\end{tabular}%
}
\end{table}

\subsection{Setup}
\label{sec:exp_setup}

\paragraph{Models \& compression format.} We compress experts of Qwen3-30B-A3B, Qwen3.5-397B-A17B~\citep{qwen2026qwen35}, and Kimi-K2.5~\citep{kimiteam2026k25} with paired $4{:}8$ NVFP4 W4A4 format (see ~\cref{app:sq_format} for format details) supported by NVIDIA's Blackwell architecture.
The fidelity of compressed checkpoint is then measured on general knowledge and reasoning tasks, with original released checkpoint (BF16 for Qwen and native INT4 for Kimi), denoted Reference, and NVIDIA's dense NVFP4 checkpoints~\citep{nvidia2025qwen330bnvfp4,nvidia2026qwen35nvfp4,nvidia2026kimik25nvfp4} provided as a baseline.
\paragraph{Compression baseline.} For joint sparsification and quantization baseline, JSQ~\citep{guo2024jsq}, SGPTQ~\citep{frantar2023sparsegpt,frantar2023gptq}, and OBR~\citep{guo2026obr} is adapted to MoE compression; \cref{app:setup} contains details on baseline implementation, compression hyperparameters.
Note that these one-shot baselines do not provide a corresponding iterative stage to which the same optimization budget can be allocated, whereas our differentiable formulation enables additional refinement from the SGPTQ solution at a higher compression cost; \cref{app:ablation} isolates the design choices within this refinement.
Broader comparison with expert compression includes 2-bit GSQ~\citep{dadgarnia2026gsq} and REAP expert pruning~\citep{lasby2025reap}, with the similar compression ratio to \ours.
\paragraph{Grouped GEMM kernels.} We compare \ours against dense grouped GEMM baselines: precision-matched NVFP4 W4A4 CuTeDSL~\citep{cutlass_blackwell_docs}, and mainstream weight-only kernels including INT4 W4A16 Marlin~\citep{frantar2025marlin} and INT2 W2A16 Humming~\citep{humming2026}.
Implementation details, kernel profiling, and serving configurations can be found in \cref{app:systems}.

\subsection{Accuracy Evaluation}
\label{sec:exp_main}

\paragraph{General knowledge.}
To evaluate general knowledge preservation, we benchmark compressed models on the OpenLLM Leaderboard v1 suite~\citep{huggingface2024openllmv1} (\cref{tab:main}).
Across all scales, \ours consistently outperforms joint sparse-quantization baselines, demonstrating that continuous reparameterization successfully shields expert representations from compound compression noise.
This resilience is particularly evident on multi-step arithmetic tasks: on Qwen3.5-397B-A17B (GSM8K), \ours achieves $71.65\%$ accuracy compared to $65.28\%$ for OBR and a near-total collapse to $4.62\%$ for JSQ, where decoupled compression disrupts expert computation.
Overall, \ours retains $93.70\%$ of reference performance on Qwen3.5-397B-A17B, while leading OBR by $5.93$ recovery points ($89.30\%$ vs. $83.37\%$) on Qwen3-30B-A3B and $5.02$ points ($96.09\%$ retention) on trillion-parameter Kimi-K2.5.

\paragraph{Reasoning.}
Next, we assess whether complex reasoning capabilities are preserved by evaluating compressed Kimi-K2.5 on AIME25~\citep{aime25}, GPQA Diamond~\citep{rein2024gpqa}, and MATH500~\citep{hendrycks2021math} under a constrained token generation budget (pass@1 averaged over $N$ trials; \cref{app:setup}).
As summarized in \cref{tab:reasoning}, \ours outperforms competing baselines across all tasks, attaining a dominant overall performance recovery of $91.44\%$.
Crucially, these benchmarks reveal that multi-step reasoning is vulnerable to compression noise, which often corrupts generation trajectories and triggers non-terminating loops that exhaust token budgets (\cref{app:jsq_termination}).
This degradation severely impacts baselines: JSQ suffers an $88.3\%$ truncation rate on AIME25, collapsing its overall performance recovery to just $17.24\%$.
In contrast, \ours maintains minimal truncation rates comparable to the uncompressed baseline.

Together, these results confirm that \ours effectively mitigates compound distortion by unifying continuous sparse support relaxation with joint weight quantization, safeguarding both factual knowledge and generation stability in large-scale MoEs.

\begin{table}[t]
\centering
\caption{Kimi-K2.5 reasoning results under paired $4{:}8$ NVFP4 W4A4 compression.
  Acc. is mean pass@1 over $N$ runs; Trunc. is the fraction exhausting the $65{,}536$-token budget, and Rec. is relative to the reference.}
\label{tab:reasoning}
\footnotesize
\setlength{\tabcolsep}{3pt}
\begin{tabular}{lcccccccc}
\toprule
& \multicolumn{2}{c}{\makecell{AIME25\\[-2pt]\footnotesize $N{=}10$}} & \multicolumn{2}{c}{\makecell{GPQA Diamond\\[-2pt]\footnotesize $N{=}5$}} & \multicolumn{2}{c}{\makecell{MATH500\\[-2pt]\footnotesize $N{=}5$}} & & \\
\cmidrule(lr){2-3} \cmidrule(lr){4-5} \cmidrule(lr){6-7}
Method & Acc. & Trunc. & Acc. & Trunc. & Acc. & Trunc. & Avg & Rec. \\
\midrule
Reference & 95.67 & 3.7 & 89.49 & 0.7 & 96.36 & 0.1 & 93.84 & 100\% \\
\midrule
JSQ       & 7.33 & 88.3 & 13.43 & 72.0 & 27.76 & 67.0 & 16.17 & 17.24\% \\
SGPTQ     & 68.33 & 15.3 & 68.48 & 2.5 & 94.24 & 1.5 & 77.02 & 82.07\% \\
OBR       & 74.67 & 12.0 & 71.62 & 1.0 & 94.64 & 0.6 & 80.31 & 85.58\% \\
Ours      & \textbf{84.67} & 6.3 & \textbf{76.46} & 0.3 & \textbf{96.28} & 0.4 & \textbf{85.80} & \textbf{91.44\%} \\
\bottomrule
\end{tabular}
\end{table}

\subsection{Efficiency Evaluation}
\begin{figure}[t]
\centering
\begin{subfigure}[t]{\dimexpr\linewidth/3\relax}%
\includegraphics[trim=0.3in 0 0.3in 0]{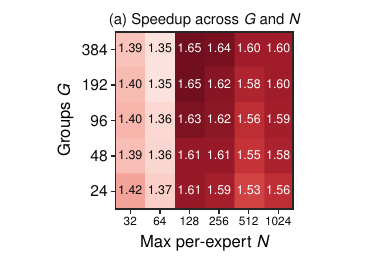}%
\phantomcaption\label{fig:gemm_speedup}%
\end{subfigure}%
\begin{subfigure}[t]{\dimexpr\linewidth/3\relax}%
\includegraphics[trim=0.3in 0 0.3in 0]{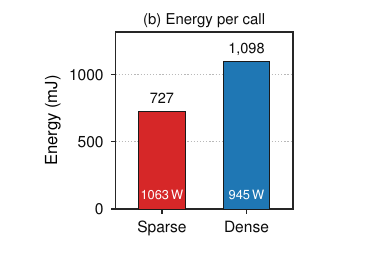}%
\phantomcaption\label{fig:gemm_energy}%
\end{subfigure}%
\begin{subfigure}[t]{\dimexpr\linewidth/3\relax}%
\includegraphics[trim=0.3in 0 0.3in 0]{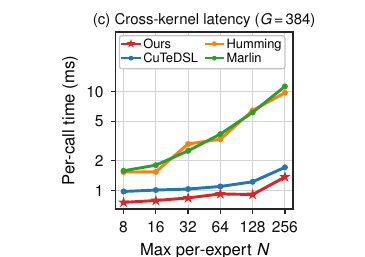}%
\phantomcaption\label{fig:gemm_xkernel}%
\end{subfigure}%
\caption{Kernel-level evaluation of the grouped sparse GEMM on the Kimi-K2.5 gate/up layer.
  (a)~Median runtime speedup of the grouped sparse GEMM kernel over dense (CuTeDSL) baseline.
  (b)~Energy to solution at iso-frequency, where the label above each bar indicates the energy per call (mJ) and the label inside each bar indicates the mean board power (W) of each kernel; measured on 1xB300.
  (c)~Median per-call latency of the deployment kernels across a range of token counts.}
\label{fig:gemm_main}
\end{figure}

\paragraph{Kernel-level efficiency.}
We first evaluate whether the custom grouped sparse GEMM kernel of \ours effectively unlocks SpTC compute density under dynamic MoE routing.
To isolate the performance impact of semi-structured sparsity, we benchmark \ours against vendor dense grouped GEMMs (CUTLASS NVFP4 CuTeDSL) using Kimi-K2.5 expert projection shapes across varying group counts $G$ and per-expert batch sizes $N$ (\cref{app:gemm_bench}).
As shown in \cref{fig:gemm_main}(a), \ours achieves a $1.35$--$1.65\times$ speedup over the dense baseline across all configurations, demonstrating robust performance across various configurations.
This acceleration directly translates into superior energy efficiency: at peak speedup, shortened execution time reduces total energy consumption from $1{,}098$ to $727$\,mJ ($1.51\times$ improvement; \cref{fig:gemm_main}(b)), showing that higher SpTC compute throughput offsets transient board power by accelerating execution to finish faster.
Finally, across the spectrum from decode ($N=8$) to prefill ($N=256$) regimes (\cref{fig:gemm_main}(c)), \ours consistently outpaces both dense FP4 and weight-only backends (INT4 Marlin, INT2 Humming).

\begin{figure}[t]
\centering
\begin{subfigure}[t]{0.25\linewidth}%
\includegraphics[trim=0.3in 0 0.3in 0]{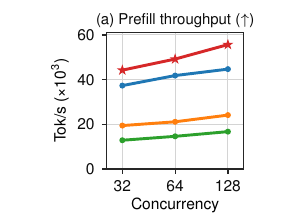}%
\phantomcaption\label{fig:e2e_prefill}%
\end{subfigure}%
\begin{subfigure}[t]{0.25\linewidth}%
\includegraphics[trim=0.3in 0 0.3in 0]{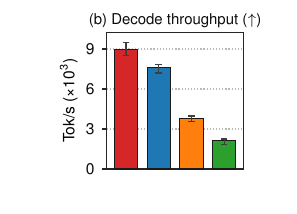}%
\phantomcaption\label{fig:e2e_decode}%
\end{subfigure}%
\begin{subfigure}[t]{0.25\linewidth}%
\includegraphics[trim=0.3in 0 0.3in 0]{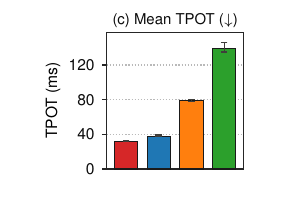}%
\phantomcaption\label{fig:e2e_tpot}%
\end{subfigure}%
\begin{subfigure}[t]{0.25\linewidth}%
\includegraphics[trim=0.3in 0 0.3in 0]{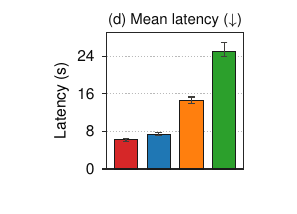}%
\phantomcaption\label{fig:e2e_e2el}%
\end{subfigure}%
\par\smallskip
\includegraphics[width=\linewidth]{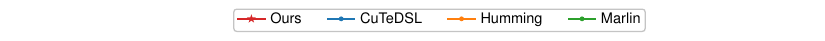}
\caption{End-to-end serving of Kimi-K2.5 on $8\times$B200 with a shared vLLM stack, varying only the expert representation and corresponding grouped-GEMM backend (\cref{tab:e2e-kimi} contains full numbers).
  (a)~Prefill throughput (total tokens/s) across concurrency, with ($4096 \ \text{in}/1 \ \text{out}$) requests.
  (b)--(d)~Decode ($384 \ \text{in} /128 \ \text{out}$) throughput (output tokens/s), mean Time Per Output Token (TPOT), and mean end-to-end request latency, with all $500$ requests in flight.
  Bars in (b)--(d) are the mean of three consecutive repeats, and whiskers show their range.}
\label{fig:e2e}
\end{figure}

\paragraph{End-to-end serving.} 
We next evaluate whether these kernel-level advantages translate into system-wide performance gains during production serving.
Using vLLM on an $8\times$ NVIDIA B200 system, we serve Kimi-K2.5 across isolated prefill and decode workloads, varying only the expert execution backend for a fair comparison (see \cref{app:e2e} for setup details).
In prefill-dominated workloads ($4096\text{ in} / 1\text{ out}$ tokens), \ours consistently yields the highest throughput across all concurrency levels, reaching $55{,}654\text{ tokens/s}$ at a concurrency of $128$---outperforming the fastest baseline (CuTeDSL) by $1.18\times$--$1.25\times$ (\cref{fig:e2e}(a)).
Under fully saturated decode workloads ($384\text{ in} / 128\text{ out}$ tokens, $500$ in-flight requests), \ours achieves $8{,}983\text{ output tokens/s}$, reducing mean Time Per Output Token (TPOT) to $32.2\text{ ms}$ and total request latency to $6.2\text{ s}$ (\cref{fig:e2e}(b)--(d)).
This translates to a $1.18\times$ boost in decode throughput along with $5.9\text{ ms}$ faster TPOT and a $1.1\text{ s}$ drop in end-to-end latency compared to dense CuTeDSL.

Together, these system- and kernel-level evaluations demonstrate that co-designing hardware-native SpTC execution with sparse low-precision compression alleviates the execution bottlenecks of trillion-parameter MoEs, delivering concurrent gains in throughput, latency, and energy efficiency.

\subsection{Expert Compression Comparison}
\label{sec:exp_compression}

Having evaluated accuracy and efficiency independently, we next compare \ours against state-of-the-art expert pruning~\citep{lasby2025reap} and dense quantization~\citep{dadgarnia2026gsq} on Kimi-K2.5 under comparable $2$--$3$-bit-per-expert weight budgets to examine the joint accuracy--memory--serving tradeoff (\cref{tab:headline}).
Within this low-bit regime, \ours recovers $96.09\%$ of reference accuracy, outperforming dense 2-bit GSQ by $1.87$ percentage points while trailing structural pruning via REAP by $1.90$ percentage points.
Importantly, \ours delivers substantial serving improvements: it achieves the highest output throughput ($8{,}983$ tok/s), lowest mean TTFT ($2{,}120$~ms), and lowest end-to-end latency ($6.2$~s), surpassing the fastest baseline by $2.36\times$ in throughput and reducing TTFT by $2.15\times$.
Although GSQ and REAP offer slightly larger KV-cache capacities due to lower expert storage footprints, both baselines yield lower throughput and higher serving latencies under the same workload and concurrency constraints.
These results indicate that reducing expert memory footprint alone does not necessarily yield efficient serving, whereas hardware-native algorithm--hardware co-design provides a favorable balance among accuracy, memory overhead, and end-to-end efficiency.

\begin{table}[t]
\centering
\caption{Comparison with alternative expert-compression methods on Kimi-K2.5.
  Serving metrics are measured with vLLM on $8\times$B200 under the same decode workload and matched concurrency ($384$ input/$128$ output tokens per request, all $500$ requests in flight) and averaged over three repeats.
  Bit counts include quantization scales and sparse metadata, KV cache is the engine-reported capacity, and bold marks the best compressed result.}
\label{tab:headline}
\small
\setlength{\tabcolsep}{3pt}
\begin{tabular}{@{}llrrrrrrr@{}}
\toprule
 & & & \multicolumn{2}{c}{OpenLLM v1} & \multicolumn{4}{c}{Serving} \\
\cmidrule(lr){4-5}\cmidrule(lr){6-9}
Method & Expert format & \makecell[r]{bits/expert\\weight} & \makecell[r]{Avg. acc.\\$\uparrow$} & \makecell[r]{Recovery\\$\uparrow$}
  & \makecell[r]{KV cache\\(tokens) $\uparrow$} & \makecell[r]{Output\\tok/s $\uparrow$} & \makecell[r]{TTFT\\(ms) $\downarrow$} & \makecell[r]{E2E\\(s) $\downarrow$} \\
\midrule
Reference & Dense INT4 W4A16  & 4.50  & 82.43 & 100\%  & 1{,}055k & 2{,}108 & 7{,}306 & 25.0 \\
\cmidrule(lr){1-9}
GSQ                  & Dense INT2 W2A16    & 2.13 & 77.66 & 94.22\% & 1{,}594k & 3{,}811 & 4{,}560 & 14.6 \\
REAP                 & $50\%$ expert pruning + INT4 & 2.25 & \textbf{80.77} & \textbf{97.99\%} & \textbf{1{,}632k} & 3{,}190 & 5{,}366 & 17.1 \\
Ours                 & Paired $4{:}8$ NVFP4 W4A4   & 2.75 & 79.21 & 96.09\% & 1{,}433k & \textbf{8{,}983} & \textbf{2{,}120} & \textbf{6.2} \\
\bottomrule
\end{tabular}
\end{table}

%% file: texts/5_conclusion.tex
\section{Conclusion}
\label{sec:conclusion}
In this work, we address the critical deployment bottleneck of Mixture-of-Experts models by exploiting hardware-accelerated Sparse Tensor Cores—an essential opportunity previously overlooked in large-scale MoE serving.
To bridge this gap, we propose \ours, an end-to-end hardware--software co-design framework that aligns hardware-native sparse low-precision representations with their efficient, routed execution.
By combining differentiable joint sparsification and quantization with a dedicated grouped sparse GEMM kernel, \ours achieves high-fidelity compression and superior serving performance at the trillion-parameter scale.
Evaluations on recent hardware demonstrate that \ours retains 96.09\% of reference accuracy while delivering up to a 1.65$\times$ kernel speedup, translating to a 1.18$\times$ throughput increase and up to a 4.03$\times$ latency reduction over standard baselines.
These findings establish that hardware--software co-design provides a highly practical paradigm to overcome memory constraints and unlock scalable MoE deployment.

\ificlrfinal
\subsection*{Artifacts}
Our code, including the grouped sparse GEMM kernel, is available on \href{https://github.com/IST-DASLab/MoE-SQ}{GitHub}.
The compressed models are released as a \href{https://huggingface.co/collections/ISTA-DASLab/moesq}{Hugging Face collection}.
\fi

\subsection*{AI use statement}
In this work, we used generative AI tools for creating scientific figures or images, editing software code, edit a paper to improve readability, format references.
We have reviewed all AI-assisted work: checking LLM-generated reference formats, code implementations of methods were verified by lead author and tested for correctness.
We take responsibility for the final content of this work, including text, claims or artifacts produced with the aid of generative AI.

%% file: texts/ack.tex
\subsubsection*{Acknowledgments}
This work was partly supported by the Institute of Information \& communications Technology Planning \& Evaluation (IITP) grant funded by the Korean government (MSIT) (No.RS-2026-25507427, Development of Efficient Architectures and Training Techniques for High-Performance Lightweight AI Models, RS-2024-00457882, National AI Research Lab Project, RS-2019-II191906, Artificial Intelligence Graduate School Program (POSTECH)),
the Korea Basic Science Institute(National research Facilities and Equipment Center) grant funded by the Korea government(MSIT) (RS-2026-25500419),
the National Research Foundation of Korea(NRF) grant funded by the Korea government(MSIT) (RS-2025-02264052, RS-2023-00210466),
and the computational results have been achieved using the Austrian Scientific Computing (ASC) infrastructure.
K.L. and D.A.'s work at ISTA DASLAB was supported in part by a generous grant from the NVIDIA Corporation.

%% file: texts/6_appendix.tex
\appendix
\crefalias{section}{appendix}
\crefalias{subsection}{subappendix}

\section{Sparse-Quantized Representation}
\label{app:sq_format}

\paragraph{Feasible set.}
Partition each row of $W$ into sparsity groups $W_{[r]}\in\mathbb{R}^{m}$ and scaling blocks $W_{[b]}\in\mathbb{R}^{d}$, and let the kernel prescribe the sparsity patterns $\mathcal{P}\subset\{0,1\}^{m}$, value grid $\mathcal{G}_w$, and scale grid $\mathcal{G}_s$.
The target feasible set in \cref{eq:feasible} is
\begin{equation}
\begin{aligned}
\mathcal{S}
&=\Big\{W:\forall r,\ \exists\rho\in\mathcal{P}\ \text{s.t.}\ W_{[r]}\odot(\mathbf{1}-\rho)=\mathbf{0}\Big\},\\
\mathcal{Q}
&=\Big\{W:\forall b,\ \exists s_b\in\mathcal{G}_s\ \text{s.t.}\ W_{[b]}\in\big(s_b\mathcal{G}_w\big)^{d}\Big\},\\
\mathcal{C}
&=\mathcal{S}\cap\mathcal{Q}.
\end{aligned}
\label{eq:feasible_detail}
\end{equation}
Because $0\in\mathcal{G}_w$, pruned positions satisfy the quantization constraint, and each shared scale governs the retained values of its block.
For Blackwell paired $4{:}8$ sparse NVFP4, $m=8$, $\mathcal{P}$ contains the six patterns that retain two adjacent pairs, $\mathcal{G}_w$ is the E2M1 grid, $\mathcal{G}_s$ contains UE4M3 scales up to a per-matrix constant, and $d=32$.
\cref{fig:p48_nvfp4} illustrates this instantiation.

\begin{center}
\begin{minipage}{0.88\linewidth}
\centering
\includegraphics[width=\linewidth]{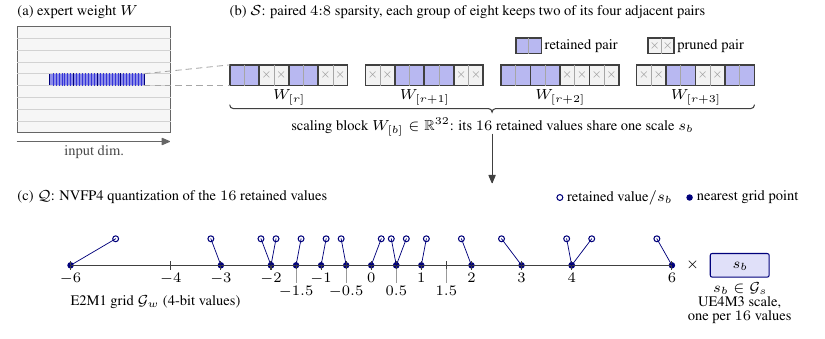}
\captionsetup{hypcap=false}
\captionof{figure}{\cref{eq:feasible_detail} instantiated for the Blackwell sparse NVFP4 instruction ($m=8$, $d=32$).
(a) An expert weight matrix; $32$ consecutive weights along the input dimension are magnified.
(b) Each group of eight retains two of its four adjacent pairs, one of the six paired $4{:}8$ patterns in $\mathcal{P}$.
(c) The $16$ retained values of a block lie on the E2M1 grid $\mathcal{G}_w=\{0,\pm0.5,\pm1,\pm1.5,\pm2,\pm3,\pm4,\pm6\}$ and share one UE4M3 (unsigned FP8) scale $s_b\in\mathcal{G}_s$.}
\label{fig:p48_nvfp4}
\end{minipage}
\end{center}

\paragraph{Storage layout.}
The packed representation stores, per sparsity group of $m=8$ weights, the four retained E2M1 values and the index of the retained pattern as sparse metadata, and one UE4M3 scale per block of $d=32$ weights:
\begin{equation}
\underbrace{4\cdot\tfrac{4}{8}}_{\text{E2M1 values}}
+\underbrace{\tfrac{4}{8}}_{\text{sparse metadata}}
+\underbrace{\tfrac{8}{32}}_{\text{UE4M3 scale}}
=2.75\ \text{bits per weight}.
\label{eq:bits_per_weight}
\end{equation}

\section{Experimental Setup}
\label{app:setup}

\paragraph{Compression scope.}
Our method and the sparse-quantized baselines compress only the routed expert FFN weights; attention, shared experts, routers, embeddings, and the LM head remain at the reference checkpoint precision.
The target representation is paired $4{:}8$ sparsity with NVFP4 weights and activations (W4A4), using $32$-element scale blocks as required by the Blackwell sparse NVFP4 path.
Details of the feasible set and storage layout are given in \cref{app:sq_format}.

\paragraph{Baselines.}
For sparse-quantized reconstruction, we adapt the official implementations of JSQ~\citep{guo2024jsq}, SGPTQ~\citep{frantar2023sparsegpt,frantar2023gptq}, and OBR~\citep{guo2026obr} to MoE experts under the same paired $4{:}8$ NVFP4 W4A4 target.
SGPTQ transfers directly, applying SparseGPT support selection followed by GPTQ with the support fixed.
The SGPTQ comparison is not compute-matched to \ours: SGPTQ uses $512$ calibration sequences for its one-shot pass, while \ours starts from the resulting SGPTQ solution and performs a further $10$ refinement epochs over $8{,}192$ sequences.
We report this difference explicitly because the comparison isolates the solution quality reached by the two optimization recipes, not their quality at equal compression cost.
For OBR, we omit the rotation step because the corresponding rotated experts cannot be executed by the grouped sparse GEMM path evaluated here; we refer to this executable adaptation simply as OBR throughout the empirical evaluation.
For the broader expert-compression comparison, GSQ~\citep{dadgarnia2026gsq} uses dense INT2 W2A16 weights, and REAP~\citep{lasby2025reap} prunes $50\%$ of routed experts while retaining the released checkpoint's INT4 weights.
These methods are compared jointly in \cref{sec:exp_compression}.

\paragraph{Calibration data.}
All compression stages use a single calibration mixture combining LLM compression calibration~\citep{neuralmagic2024llmcompressioncalibration}, OpenThoughts-114k~\citep{guha2025openthoughtsdatarecipesreasoning}, and FineWeb-Edu~\citep{lozhkov2024fineweb-edu} in a $10{:}45{:}45$ ratio.
The mixture is chosen to cover a broad range of data, spanning curated calibration text, long-form reasoning traces, and filtered web text.
Documents are packed into sequences of $4{,}096$ tokens.
SGPTQ initialization draws $512$ sequences, and support--value refinement draws $8{,}192$ sequences, with a further $128$ held out for validation.
The same mixture is used for every model and for every reported arm.

\paragraph{Compression hyperparameters.}
\cref{tab:hparams} gives the per-model optimization configuration of the router-weighted objective used for the main results.

\begin{table}[h]
\centering
\caption{Per-model compression hyperparameters.
  Learning rates are for the support logits $A$ and latent weights $V$; the Gumbel temperature $\tau_t$ of \cref{eq:gumbel} is annealed over the schedule shown.}
\label{tab:hparams}
\small
\begin{tabular}{@{}lccc@{}}
\toprule
 & Kimi-K2.5 & Qwen3.5-397B-A17B & Qwen3-30B-A3B \\
\midrule
MoE blocks compressed        & 60    & 60    & 48 \\
Routed experts per block     & 384   & 512   & 128 \\
\addlinespace
Calibration sequences        & 8{,}192 & 8{,}192 & 8{,}192 \\
Batch size (micro-batch)     & 64 (8)  & 64 (8)  & 64 (16) \\
Epochs per block             & 10      & 10      & 10 \\
\addlinespace
Optimizer                    & \multicolumn{3}{c}{Lion~\citep{chen2023lion}} \\
Support-logit learning rate ($A$) & $2{\times}10^{-4}$ & $2{\times}10^{-4}$ & $2.5{\times}10^{-4}$ \\
Latent-weight learning rate ($V$) & $5{\times}10^{-5}$ & $2.5{\times}10^{-5}$ & $5{\times}10^{-5}$ \\
Gumbel temperature $\tau_t$  & \multicolumn{3}{c}{$2.0 \rightarrow 0.05$} \\
SGPTQ damping (percdamp)     & 0.1   & 0.1   & 0.01 \\
\addlinespace
Devices                      & $8\times$B200 & $8\times$B200 & $4\times$B200 \\
Wall-clock                   & ${\approx}27.6$\,h & ${\approx}18.9$\,h & ${\approx}13$\,h \\
\bottomrule
\end{tabular}
\end{table}

\paragraph{Evaluation protocol.}
All accuracy evaluations are served through vLLM with the KV-cache dtype fixed within each model family.
We evaluate OpenLLM Leaderboard v1~\citep{huggingface2024openllmv1} with \texttt{lm-eval-harness}~\citep{eval-harness} on all three models.
For reasoning, we evaluate Kimi-K2.5 on AIME25~\citep{aime25}, GPQA Diamond~\citep{rein2024gpqa}, and MATH500~\citep{hendrycks2021math} following the GSQ protocol~\citep{dadgarnia2026gsq}.
Pass@1 is averaged over $10$, $5$, and $5$ runs, respectively, with temperature $1.0$, top-$p$ $0.95$, and a $65{,}536$-token generation budget.

\section{Additional Accuracy Results}
\label{app:accuracy}

\subsection{Ablations}
\label{app:ablation}

SGPTQ is a one-shot procedure and does not provide an iterative stage to which additional optimization budget can be allocated, whereas our differentiable formulation enables its sparse-quantized solution to serve as the initialization for further refinement at a higher compression cost.
We hold this added budget fixed and ablate the two design choices enabled by the refinement stage on Qwen3-30B-A3B: the router-weighted objective and joint support--value adaptation.
All arms use the same SGPTQ initialization, calibration setup, and $10$-epoch refinement budget, and the full \ours configuration is the Qwen3-30B-A3B column of \cref{tab:hparams}, with per-task results in \cref{tab:ablation}.

\paragraph{Router-weighted reconstruction objective.}
Replacing the uniform special case $w_{l,e}=\mathbf{1}$ with the router-weighted objective of \cref{eq:diag} increases the OpenLLM average from $64.68$ to $65.44$ and recovery from $88.26\%$ to $89.30\%$, with improvements on all six tasks.
We tune the learning rates separately for the uniform and router-weighted objectives rather than forcing matched rates across objectives with different gradient scales.
The selected support-logit and latent-weight learning rates are $5\times10^{-4}$ and $1\times10^{-4}$ for the uniform objective, and $2.5\times10^{-4}$ and $5\times10^{-5}$ for the router-weighted objective, respectively.

\paragraph{Support--value adaptation.}
With the initial SGPTQ support fixed, adapting only the latent weights $V$ reaches $64.44$ average accuracy, whereas jointly adapting the support logits $A$ and latent weights $V$ under the router-weighted objective reaches $65.44$, an improvement of $1.00$ percentage point.

\begin{table}[h]
\centering
\caption{Ablations on Qwen3-30B-A3B under the paired $4{:}8$ NVFP4 W4A4 target.
  $\mathcal{L}^{\mathrm{unif}}$ denotes \cref{eq:diag} with $w_{l,e}=\mathbf{1}$, and ``$V$ only'' keeps the initial SGPTQ support fixed.
  Avg. is the unweighted mean across the six OpenLLM tasks, Rec. is relative to the BF16 reference in \cref{tab:main}, and bold marks the best result in each column.}
\label{tab:ablation}
\small
\setlength{\tabcolsep}{4pt}
\resizebox{\linewidth}{!}{%
\begin{tabular}{llcccccccc}
\toprule
Objective & Updated & ARC-c & GSM8K & MMLU & WG & HSwag & TQA & Avg & Rec. \\
\midrule
$\mathcal{L}^{\mathrm{unif}}$ & $(A,V)$ & 59.30 & 77.18 & 69.84 & 66.93 & 63.25 & 51.55 & 64.68 & 88.26\% \\
$\mathcal{L}^{\mathrm{rw}}$ & $V$ only & 57.42 & 79.15 & \textbf{70.18} & 66.38 & 63.41 & 50.08 & 64.44 & 87.95\% \\
\midrule
$\mathcal{L}^{\mathrm{rw}}$ (\ours) & $(A,V)$ & \textbf{59.73} & \textbf{79.45} & 69.89 & \textbf{67.80} & \textbf{63.60} & \textbf{52.15} & \textbf{65.44} & \textbf{89.30\%} \\
\bottomrule
\end{tabular}%
}
\end{table}

\subsection{Per-Task Expert-Compression Comparison}
\label{app:other_families}

This section provides the per-task accuracy results underlying the expert-compression comparison of \cref{sec:exp_compression} and \cref{tab:headline}.
Unlike \cref{tab:main,tab:reasoning}, which compare methods under the same sparse W4A4 target, the methods here use different compression mechanisms and deployment formats.

\begin{table}[h]
\centering
\caption{OpenLLM Leaderboard v1 comparison with alternative expert-compression methods on Kimi-K2.5.
  GSQ uses dense INT2 W2A16 weights, REAP prunes $50\%$ of routed experts while retaining the released INT4 weights, and \ours uses paired $4{:}8$ NVFP4 W4A4.
  Recovery is relative to the reference row, and bold marks the best compressed result.}
\label{tab:other_families}
\small
\setlength{\tabcolsep}{4pt}
\resizebox{\linewidth}{!}{%
\begin{tabular}{llcccccccc}
\toprule
Method & Expert format & ARC-c & GSM8K & MMLU & WG & HSwag & TQA & Avg & Rec. \\
\midrule
Reference     & Dense INT4 W4A16   & 74.06 & 94.39 & 89.53 & 81.93 & 91.96 & 62.69 & 82.43 & 100\% \\
\cmidrule(lr){1-10}
GSQ           & Dense INT2 W2A16   & 64.51 & 91.96 & 84.35 & 80.90 & 86.01 & 58.24 & 77.66 & 94.22\% \\
REAP          & $50\%$ expert pruning + INT4 & \textbf{72.35} & \textbf{93.71} & \textbf{87.72} & 82.00 & \textbf{89.25} & 59.60 & \textbf{80.77} & \textbf{97.99\%} \\
\cmidrule(lr){1-10}
Ours          & Paired $4{:}8$ NVFP4 W4A4  & 69.97 & 88.70 & 85.34 & \textbf{82.40} & 88.57 & \textbf{60.28} & 79.21 & 96.09\% \\

\bottomrule
\end{tabular}%
}
\end{table}

\subsection{Non-Terminating Generations under JSQ}
\label{app:jsq_termination}

The reasoning degradation of JSQ in \cref{tab:reasoning} is driven largely by non-terminating generations rather than by ordinary answer errors.
Under the same paired $4{:}8$ NVFP4 W4A4 target, JSQ often enters increasingly repetitive reasoning trajectories that continue until the $65{,}536$-token budget is exhausted, whereas OBR and \ours typically remain substantially more concise and terminate normally.
\cref{fig:reasoning_traces} illustrates this failure mode on representative prompts.

\begin{figure}[h]
\centering
\begin{minipage}{\linewidth}
\footnotesize
\textbf{(a) MATH500, problem 457.} \textit{``Determine the remainder of 54 (mod 6).''}

\smallskip
\textbf{\ours} \textnormal{(281 characters, terminated, correct)}
\begin{Verbatim}
To determine the remainder of 54 when divided by 6, we can perform the division:

$$54 \div 6 = 9 \text{ with remainder } 0$$

This means:
$$54 = 6 \times 9 + 0$$

Therefore, the remainder is 0.

ANSWER: $0$
\end{Verbatim}

\textbf{JSQ} \textnormal{(260{,}474 characters, unterminated; opening, then the final 260 characters)}
\begin{Verbatim}
We need to determine the remainder of 54 mod 6, which means the remainder when 54
is divided by 6.

Recall that "mod 6" means dividing by 6 and finding the remainder. The remainder
when dividing by 6 is the leftover after the integer division.

Let's perform the division: [...]

The remainder is 0 because 54 is exactly divisible by 6 [...]
\end{Verbatim}
\begin{Verbatim}
 the the the the the the the the the the the the the the the the the the the the
the the the the the the the the the the the the the the the the the the the the
the the the the the the the the the the the the the the the the the the the the
\end{Verbatim}

\smallskip
\textbf{(b) AIME25, problem 14.} All three methods exhaust the budget on this problem; final characters of each generation:

\smallskip
\textbf{\ours} \textnormal{(174{,}615 characters, unterminated)}
\begin{Verbatim}
However, the uniform distribution assumption is likely correct because H is a
"random" subset in the additive sense, or due to the group action.

Given the symmetry, it is plausible that the number of pairs summing to any
particular d depends only on the valuation of d.

Thus, for d with v_3 = 3, N
\end{Verbatim}

\textbf{OBR} \textnormal{(175{,}294 characters, unterminated)}
\begin{Verbatim}
Alternatively, note that H is the set of cubes. [...] Then the sum is
\end{Verbatim}

\textbf{JSQ} \textnormal{(194{,}373 characters, unterminated)}
\begin{Verbatim}
 ... ... ... ... ... ... ... ... ... ... ... ... ... ... ... ... ... ... ... ...
... ... ... ... ... ... ... ... ... ... ... ... ... ... ... ... ... ... ... ...
... ... ... ... ... ... ... ... ... ... ... ... ... ... ... ... ... ... ... ...
\end{Verbatim}
\end{minipage}
\caption{Representative reasoning trajectories under the same paired $4{:}8$ NVFP4 W4A4 target.
  In (a), \ours terminates with the correct answer, whereas JSQ reaches the same answer early but continues into a degenerate repetitive tail until the generation budget is exhausted.
  In (b), all methods exhaust the budget on the same difficult prompt, but \ours and OBR remain in coherent reasoning while JSQ has collapsed into repetition.
  Text is verbatim from the evaluation records, with \texttt{[...]} marking elisions and line breaks inserted for readability.}
\label{fig:reasoning_traces}
\end{figure}

To quantify the effect, \cref{tab:jsq_termination} separates answer quality from termination behavior.
Accuracy on terminated generations measures how often a method answers correctly once it does terminate, while the matched-subset comparison restricts all methods to problems for which JSQ terminates at least once.
This avoids attributing the full gap solely to the generation cutoff.

\begin{table}[h]
\centering
\caption{Termination behavior of sparse-quantized Kimi-K2.5 models.
  Accuracy on terminated generations isolates answer quality after successful termination, while the matched-subset comparison controls for JSQ terminating on substantially fewer problems.
  Coverage is the number of problems for which JSQ terminates at least once out of the benchmark total.}
\label{tab:jsq_termination}
\small
\setlength{\tabcolsep}{4pt}
\resizebox{\linewidth}{!}{%
\begin{tabular}{lccccccccc}
\toprule
& \multicolumn{3}{c}{Accuracy on terminated generations} & \multicolumn{4}{c}{Matched subset} & \multicolumn{2}{c}{Median terminated (chars)} \\
\cmidrule(lr){2-4} \cmidrule(lr){5-8} \cmidrule(lr){9-10}
Benchmark & Ours & OBR & JSQ & Coverage & Ours & OBR & JSQ & Ours & JSQ \\
\midrule
AIME25        & 90.39 & 84.85 & 62.86 & 11/30   & 100.00 & 100.00 & 20.00 & 42{,}580 & 83{,}504 \\
GPQA Diamond  & 76.70 & 72.35 & 48.01 & 123/198 & 75.12  & 71.06  & 21.63 & 43{,}810 & 105{,}104 \\
MATH500       & 96.63 & 95.21 & 84.12 & 379/500 & 96.89  & 95.41  & 36.62 & 5{,}426  & 34{,}486 \\
\bottomrule
\end{tabular}%
}
\end{table}

The qualitative and quantitative results tell the same story.
JSQ is heavily truncated on all three benchmarks, and even among generations that terminate, its accuracy remains substantially below OBR and \ours.
Restricting evaluation to the matched subset does not close the gap.
Thus, under this aggressive sparse-quantized target, poor compression can degrade long-horizon generation dynamics, producing rambling or repetitive trajectories that fail to terminate rather than simply increasing the frequency of conventional wrong answers~\citep{lucas2025reasoningpruning}.
We restrict this observation to the evaluated JSQ adaptation and do not claim that non-termination is universal to JSQ or sparse quantization in general.

\section{Grouped Sparse GEMM Kernel and Serving}
\label{app:systems}

\subsection{Implementation Details}
\label{app:kernel_impl}

\paragraph{Operand layout and scheduling.}
Weights are packed offline into E2M1 values, sparse metadata, and one UE4M3 scale per block of $32$ logical weights, as detailed in \cref{app:sq_format}.
Activations are quantized at run time into the corresponding $32$-element scale blocks by $\mathcal{Q}_a$ of \cref{eq:diag}.
We store activations and outputs in a batched layout with a fixed per-expert capacity $\bar N$, so every operand of expert $e$ lies at a fixed stride and no pointer arrays are needed.
The token counts remain on the device, and the persistent scheduler assigns expert $e$ exactly $\lceil N_e/T_N\rceil$ tiles along the token dimension; only its last partial tile can touch padded rows, whose outputs are discarded downstream.

\paragraph{Tile configurations.}
The MMA tile is $T_M\times T_N\times 256$, with $T_M\in\{128,256\}$ in one- or two-SM mode and $T_N\in\{128,256\}$~\citep{cutlass_blackwell_docs}.
The UE4M3 block scales of both operands are applied inside the MMA, while their per-tensor scales are folded into one factor $\alpha_e$ per expert that the epilogue applies to the FP32 accumulator before writing BF16.
The four tile shapes and admissible cluster shapes produce $30$ kernel variants.
At startup, the serving backend times these variants once per expert GEMM geometry on the loaded weights and freezes the selected variant before CUDA-graph capture.

\paragraph{Serving integration.}
We expose the kernel to vLLM~\citep{kwon2023vllm} through the same fused-MoE interface as the dense NVFP4 backend built on FlashInfer~\citep{ye2025flashinfer}.
The sparse backend implements fused dispatch--scatter with activation quantization, the two grouped GEMMs, and fused SwiGLU with requantization, while routing, dispatch planning, batched preparation and finalization, and token combination remain shared.
Unlike the dense NVFP4 path, which uses $16$-element scale blocks, our activation-quantization kernel forms $32$-element blocks and emits the scale layout required by the sparse MMA.
We fuse this quantization with token dispatch and scatter for the first expert GEMM and with SwiGLU output preparation for the second.

\subsection{Kernel Benchmark}
\label{app:gemm_bench}

We benchmark our grouped sparse NVFP4 GEMM on a single B200 against the CUTLASS dense NVFP4 grouped GEMM, which we run in two operand layouts.
\cref{tab:gemm_bench_config} summarizes the layer shapes and sweep ranges.
Per-expert token counts follow the deterministic imbalanced pattern $\{N,3N/4,N/2,N/4\}$ repeated across groups.

\begin{figure}[t]
\centering
\includegraphics[width=\linewidth]{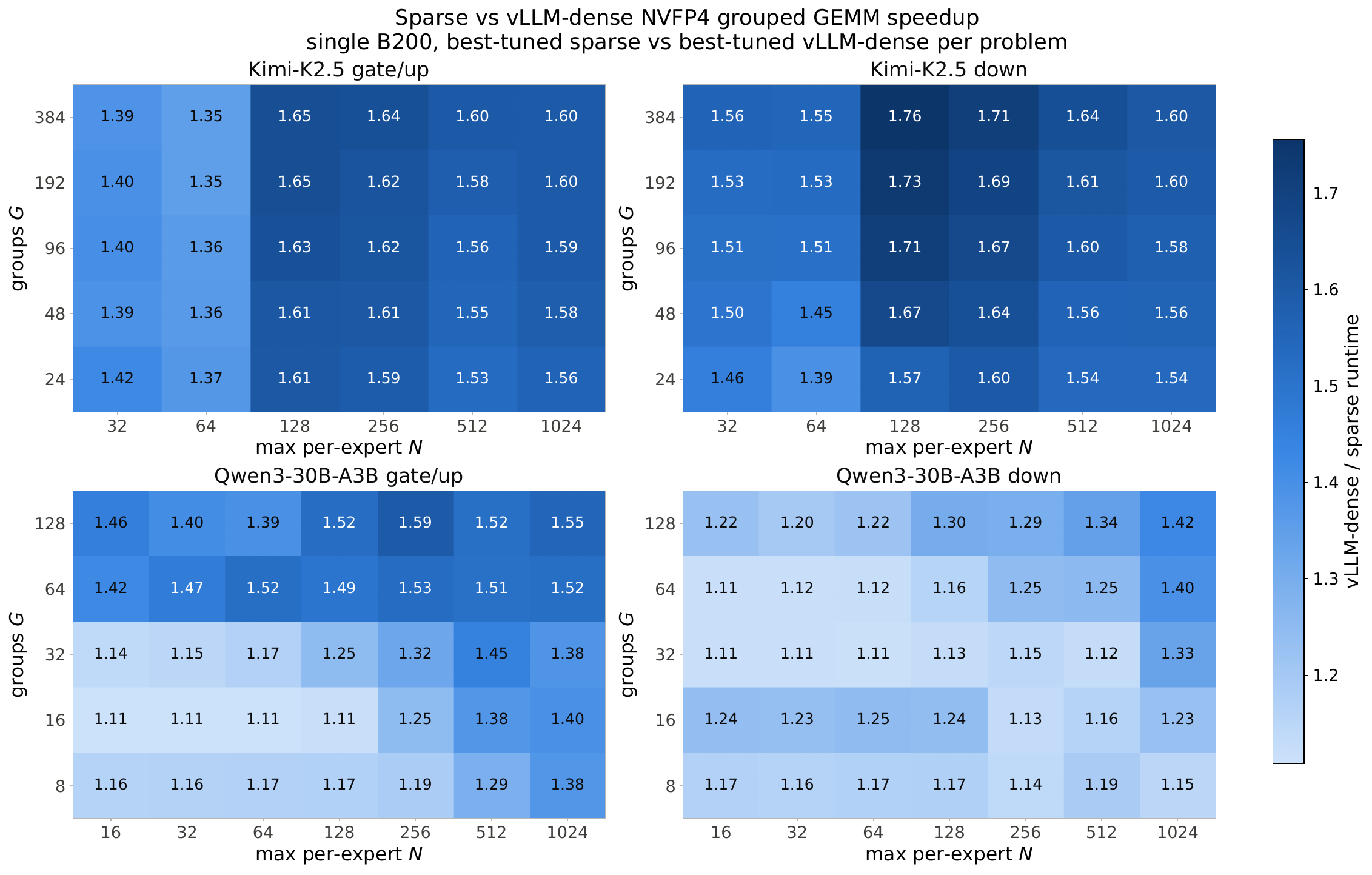}
\caption{Autotuned speedup of the grouped sparse GEMM over the dense NVFP4 grouped GEMM (native layout) across the four benchmarked MoE layers on B200.
  Rows are expert-group count $G$ and columns are maximum per-expert token count $N$.}
\label{fig:gemm_speedup_grid}
\end{figure}

\paragraph{Operand layout.}
\cref{sec:kernel} describes the operand orientation of the sparse kernel.
The dense baseline, Dense (native), keeps the activation-stationary layout used by vLLM; the speedups in \cref{fig:gemm_main}(a) and \cref{fig:gemm_speedup_grid} are relative to it.
The dense control, Dense (transposed), runs the same kernel in the orientation of the sparse kernel, so that it differs from our kernel only in sparsity.

\begin{table}[h]
\centering
\small
\caption{Grouped-GEMM benchmark shapes.
  $(M,K)$ are the output and contraction dimensions; $N$ is the maximum per-expert token count and $G$ the expert-group count.}
\label{tab:gemm_bench_config}
\begin{tabular}{llccc}
\toprule
Model & Layer & $(M, K)$ & $N$ sweep & $G$ sweep \\
\midrule
\multirow{2}{*}{Kimi-K2.5} & gate/up & $(4096, 7168)$ & \multirow{2}{*}{$\{32,64,128,256,512,1024\}$} & \multirow{2}{*}{$\{24,48,96,192,384\}$} \\
                          & down    & $(7168, 2048)$ & & \\
\midrule
\multirow{2}{*}{Qwen3-30B-A3B} & gate/up & $(1536, 2048)$ & \multirow{2}{*}{$\{16,32,64,128,256,512,1024\}$} & \multirow{2}{*}{$\{8,16,32,64,128\}$} \\
                               & down    & $(2048, 768)$  & & \\
\bottomrule
\end{tabular}
\end{table}

\paragraph{Timing and autotuning.}
We time pure on-device kernel execution with CUDA-graph replay following QUTLASS~\citep{qutlass2025}; allocation, operand preparation, sparse compression, and kernel initialization are excluded.
We report median replay latency.
Each problem/kernel pair is autotuned independently over SM mode, MMA tile shape, and Blackwell cluster configuration (\cref{tab:gemm_search}), and the fastest valid variant is selected.
Effective throughput and speedup are
\[
\mathrm{TFLOP/s}_{\mathrm{eff}} = \frac{2 M K \sum_{g=1}^{G} N_g}{t \cdot 10^{12}},
\qquad
\mathrm{speedup} = \frac{t_{\mathrm{dense}}}{t_{\mathrm{sparse}}}.
\]

\begin{table}[h]
\centering
\small
\caption{Autotuner search space.
  Tile-$M$ is fixed by the SM mode and tile-$K{=}256$ throughout; supported tile-$N$ values differ across kernels.}
\label{tab:gemm_search}
\begin{tabular}{lccc}
\toprule
Kernel & SM mode (tile-$M$) & tile-$N$ & tile-$K$ \\
\midrule
Sparse (ours)           & 1-SM ($128$), 2-SM ($256$) & $\{128, 256\}$      & $256$ \\
Dense (native)          & 1-SM ($128$), 2-SM ($256$) & $\{128, 192, 256\}$ & $256$ \\
Dense (transposed)      & 1-SM ($128$), 2-SM ($256$) & $\{128, 192\}$      & $256$ \\
\bottomrule
\end{tabular}
\end{table}

\paragraph{Speedup across shapes.}
Across the full $G\times N$ sweep, the sparse kernel is faster in every configuration.
On Kimi-K2.5, geometric-mean speedups are $1.52\times$ for gate/up and $1.58\times$ for down; on Qwen3-30B-A3B they are $1.33\times$ and $1.20\times$, respectively.
The gain depends primarily on per-expert token count and contraction size: it is smallest for short, memory-dominated problems and approaches the sparse-MMA advantage as arithmetic intensity increases.
\cref{fig:gemm_speedup_grid} gives the full sweep.

\paragraph{Energy measurement.}
We measure whole-board energy with NVML over sustained CUDA-graph replay on an otherwise idle, exclusive B300, using $200$ warmup iterations and a pinned SM clock for the iso-frequency comparison.
We report energy per call and average board power.
Because the sparse kernel draws more instantaneous power but finishes sooner, energy-to-solution is the relevant metric; \cref{tab:gemm_energy} reports both iso-frequency and power-capped measurements.

\begin{table}[h]
\centering
\small
\caption{Energy-to-solution for the Kimi-K2.5 gate/up grouped GEMM ($M{=}4096$, $N{=}128$, $K{=}7168$, $G{=}384$) on B300.
  Iso-frequency is the primary comparison; the power-capped regime is included for completeness.
  Dense columns are the two operand layouts of the same CUTLASS kernel (\cref{app:gemm_bench}); speedup and energy reduction are those of the sparse kernel over each dense column.}
\label{tab:gemm_energy}
\resizebox{\linewidth}{!}{%
\begin{tabular}{l rrr rrr}
\toprule
 & \multicolumn{3}{c}{Iso-frequency ($1702$\,MHz)} & \multicolumn{3}{c}{Power-capped ($\approx 1095$\,W)} \\
\cmidrule(lr){2-4} \cmidrule(lr){5-7}
 & Sparse & \makecell[r]{Dense\\(transposed)} & \makecell[r]{Dense\\(native)} & Sparse & \makecell[r]{Dense\\(transposed)} & \makecell[r]{Dense\\(native)} \\
\midrule
Avg.\ power (W)        & $1063$  & $865$   & $945$   & $1095$  & $1096$  & $1092$  \\
SM clock (MHz)         & $1702$  & $1702$  & $1702$  & $1725$  & $1897$  & $1867$  \\
Latency (ms)           & $0.687$ & $1.348$ & $1.164$ & $0.679$ & $1.183$ & $1.131$ \\
Throughput (TFLOP/s)   & $2627$  & $1339$  & $1550$  & $2657$  & $1525$  & $1595$  \\
Energy/call (mJ)       & $727$   & $1166$  & $1098$  & $745$   & $1298$  & $1237$  \\
Efficiency (TFLOP/J)   & $2.48$  & $1.55$  & $1.64$  & $2.42$  & $1.39$  & $1.46$  \\
\midrule
Speedup            & --  & $1.96\times$ & $1.69\times$ & --  & $1.74\times$ & $1.67\times$ \\
Energy reduction   & --  & $1.60\times$ & $1.51\times$ & --  & $1.74\times$ & $1.66\times$ \\
\bottomrule
\end{tabular}%
}
\end{table}

\paragraph{Cross-kernel comparison.}
\cref{fig:gemm_main}(c) compares the deployment kernels used in \cref{fig:e2e} on the Kimi-K2.5 gate/up layer at $G{=}384$, spanning decode-like to prefill-like token counts.
Timings include activation quantization for the W4A4 kernels and dequantization for the W4A16/W2A16 kernels, from BF16 input activations to BF16 output.
Our sparse kernel is fastest across the full range, including against INT4 Marlin and INT2 Humming; \cref{tab:xkernel} reports the measured latencies.

\begin{table}[h]
\centering
\small
\caption{Cross-kernel grouped-GEMM latency ($\mu$s) on the Kimi-K2.5 gate/up layer ($M{=}4096$, $K{=}7168$, $G{=}384$) on B200.
  Timings include activation quantization/dequantization and use the same ragged token pattern as the main sweep.}
\label{tab:xkernel}
\begin{tabular}{@{}lrrrrrrr@{}}
\toprule
 & bits/p & $N{=}8$ & $N{=}16$ & $N{=}32$ & $N{=}64$ & $N{=}128$ & $N{=}256$ \\
\midrule
Ours (paired $4{:}8$ W4A4) & 2.750 & \textbf{760} & \textbf{793} & \textbf{843} & \textbf{925} & \textbf{912} & \textbf{1{,}370} \\
NVFP4-W4A4 CuTeDSL    & 4.500 & 979 & 1{,}012 & 1{,}036 & 1{,}100 & 1{,}226 & 1{,}717 \\
INT4-W4A16 Marlin     & 4.500 & 1{,}584 & 1{,}805 & 2{,}519 & 3{,}717 & 6{,}163 & 11{,}272 \\
INT2-W2A16 Humming    & 2.125 & 1{,}538 & 1{,}545 & 2{,}959 & 3{,}299 & 6{,}446 & 9{,}733 \\
\bottomrule
\end{tabular}
\end{table}

\subsection{End-to-End Serving Benchmark}
\label{app:e2e}

We serve Kimi-K2.5 on $8\times$B200 with vLLM under a shared DP${=}8$ $\times$ EP topology using DeepEP low-latency all-to-all.
Across all backends, routing, dispatch planning, batched preparation and finalization, batching, memory controls, token combination, and output layout are held fixed; only the expert representation and its associated compute path differ.
The number of concurrent requests is pinned with \texttt{-{}-max-concurrency}, so every backend is measured at matched concurrency.
\cref{tab:e2e-setup} gives the shared configuration and \cref{tab:e2e-kimi} the full results corresponding to \cref{fig:e2e}.

\paragraph{Choice of comparators.}
We compare our grouped sparse GEMM (paired $4{:}8$ NVFP4 W4A4) with dense NVFP4 W4A4 CuTeDSL, INT4 W4A16 Marlin, and INT2 W2A16 Humming.
We use the batched expert variants of all kernels so that they share the same dispatch path and output layout.

\paragraph{Workloads.}
We report two workloads, each of which isolates one serving phase, and every measurement issues $500$ requests.
The prefill workload uses $4096$ input tokens and $1$ output token per request at concurrencies of $32$, $64$, and $128$, where concurrency denotes the number of concurrent requests summed over the eight DP engines.
Since each request emits a single token, we report total token throughput for this workload.
The decode workload uses $384$ input and $128$ output tokens per request with a concurrency cap of at least $500$, so that all $500$ requests are in flight.
At $512$ tokens per request, this requires $256$K KV-cache tokens, well below the capacity of every backend in \cref{tab:e2e-kimi}, so no backend is limited by KV-cache capacity.
The decode workload is measured three consecutive times within the same session, with concurrency caps of $512$, $1024$, and $2048$, all of which exceed the number of issued requests and therefore yield the same operating point.
Throughput varies by up to $22\%$ across these repeats, generally increasing as the server warms up, so we report the mean of the three repeats together with their range.
The first measurement of each session is excluded for all backends, since it absorbs server warmup.

\begin{table}[h]
\centering
\caption{End-to-end serving on Kimi-K2.5 ($8\times$ B200, matched concurrency); the results of \cref{fig:e2e} in full.}
\label{tab:e2e}
\small

\begin{subtable}[t]{\linewidth}
\centering
\caption{Serving configuration, shared by all four backends; only the expert kernel and weight format differ.}
\label{tab:e2e-setup}
\begin{tabular}{@{}ll@{}}
\toprule
Hardware       & $8\times$ NVIDIA B200 \\
Engine         & vLLM, DP${=}8$ $\times$ EP, DeepEP low-latency all-to-all \\
Fixed controls & gpu-mem-util $0.97$, max-model-len $4608$, max-batched-tokens $256$ \\
KV cache dtype & \texttt{bfloat16}, pinned for all backends \\
Prefill workload & $4096$ in / $1$ out, concurrency $\in\{32,64,128\}$, $500$ requests \\
Decode workload  & $384$ in / $128$ out, all $500$ requests in flight, $3$ repeats \\
\bottomrule
\end{tabular}
\end{subtable}

\vspace{1em}

\begin{subtable}[t]{\linewidth}
\centering
\caption{Full serving results corresponding to \cref{fig:e2e}.
  Best per row in \textbf{bold}.
  Prefill rows are measured at each concurrency of the prefill workload.
  Decode and latency rows are the mean over three repeats of the decode workload with all $500$ requests in flight, and the range row gives the minimum and maximum decode throughput across the repeats.
  $^{\ast}$Maximum concurrency is reported by vLLM at engine start relative to a full-length request.
  $^{\dagger}$Prefill emits one token per request, so prefill rows report total token throughput; decode rows report output token throughput.}
\label{tab:e2e-kimi}
\begin{tabular}{@{}lrrrr@{}}
\toprule
  & \makecell[r]{\textbf{Ours}\\(Paired $4{:}8$ W4A4)}
  & \makecell[r]{NVFP4-W4A4\\CuTeDSL}
  & \makecell[r]{INT4-W4A16\\Marlin}
  & \makecell[r]{INT2-W2A16\\Humming} \\
\midrule
\multicolumn{5}{@{}l}{\emph{Throughput} ($\uparrow$)} \\
\quad Prefill, concurrency $32$ (tok/s)$^{\dagger}$  & \textbf{44{,}223} & 37{,}314 & 12{,}864 & 19{,}430 \\
\quad Prefill, concurrency $64$ (tok/s)$^{\dagger}$  & \textbf{49{,}184} & 41{,}802 & 14{,}615 & 21{,}120 \\
\quad Prefill, concurrency $128$ (tok/s)$^{\dagger}$ & \textbf{55{,}654} & 44{,}678 & 16{,}715 & 24{,}116 \\
\quad Decode (tok/s)                                & \textbf{8{,}983}  & 7{,}619  & 2{,}108  & 3{,}811 \\
\quad Decode, range (tok/s)                         & 8{,}494--9{,}512  & 7{,}213--7{,}831 & 1{,}843--2{,}252 & 3{,}586--3{,}985 \\
\addlinespace
\multicolumn{5}{@{}l}{\emph{Latency}, decode workload ($\downarrow$)} \\
\quad End-to-end mean (s)               & \textbf{6.2}   & 7.3    & 25.0  & 14.6 \\
\quad End-to-end p$99$ (s)              & \textbf{7.0}   & 8.3    & 30.1  & 16.6 \\
\quad TPOT mean (ms)                    & \textbf{32.2}  & 38.1   & 139.3 & 79.0 \\
\quad TPOT median (ms)                  & \textbf{32.8}  & 38.8   & 140.3 & 80.2 \\
\quad TTFT mean (ms)                    & \textbf{2{,}120} & 2{,}505 & 7{,}306 & 4{,}560 \\
\addlinespace
\multicolumn{5}{@{}l}{\emph{Capacity} ($\uparrow$)} \\
\quad KV cache (tokens)                 & 1{,}432{,}608 & 1{,}098{,}720 & 1{,}054{,}848 & \textbf{1{,}593{,}760} \\
\quad Max concurrency$^{\ast}$          & 310.9$\times$ & 238.4$\times$ & 228.9$\times$ & \textbf{345.9$\times$} \\
\bottomrule
\end{tabular}
\end{subtable}
\end{table}

%% file: reference.bib
@misc{eval-harness,  
  author       = {Sutawika, Lintang and
                  Schoelkopf, Hailey and
                  Gao, Leo and
                  Abbasi, Baber and
                  Biderman, Stella and
                  Tow, Jonathan and
                  fattori, ben and
                  Lovering, Charles and
                  farzanehnakhaee70 and
                  Phang, Jason and
                  Thite, Anish and
                  Fazz and
                  Aflah and
                  Muennighoff, Niklas and
                  Wang, Thomas and
                  sdtblck and
                  nopperl and
                  gakada and
                  tttyuntian and
                  researcher2 and
                  Etxaniz, Julen and
                  Chris and
                  Lee, Hanwool Albert and
                  Kasner, Zdeněk and
                  Khalid and
                  LSinev and
                  Hsu, Jeffrey and
                  Kanekar, Anjor and
                  KonradSzafer and
                  AndyZwei},
  title        = {EleutherAI/lm-evaluation-harness: v0.4.3},
  month        = jul,
  year         = 2024,
  publisher    = {Zenodo},
  version      = {v0.4.3},
  doi          = {10.5281/zenodo.12608602},
  url          = {https://doi.org/10.5281/zenodo.12608602},
}

@misc{huggingface2024openllmv1,
  author       = {{Hugging Face}},
  title        = {Open {LLM} Leaderboard v1},
  year         = {2024},
  howpublished = {Hugging Face Leaderboards documentation},
  url          = {https://huggingface.co/docs/leaderboards/en/open_llm_leaderboard/archive}
}

@inproceedings{jang2017gumbel,
  title     = {Categorical Reparameterization with {Gumbel-Softmax}},
  author    = {Jang, Eric and Gu, Shixiang and Poole, Ben},
  booktitle = {International Conference on Learning Representations (ICLR)},
  year      = {2017},
  url       = {https://arxiv.org/abs/1611.01144}
}

@inproceedings{maddison2017concrete,
  title     = {The Concrete Distribution: A Continuous Relaxation of Discrete Random Variables},
  author    = {Maddison, Chris J. and Mnih, Andriy and Teh, Yee Whye},
  booktitle = {International Conference on Learning Representations (ICLR)},
  year      = {2017},
  url       = {https://arxiv.org/abs/1611.00712}
}

@article{bengio2013ste,
  title   = {Estimating or Propagating Gradients Through Stochastic Neurons for Conditional Computation},
  author  = {Bengio, Yoshua and L{\'e}onard, Nicholas and Courville, Aaron},
  journal = {arXiv preprint arXiv:1308.3432},
  year    = {2013},
  url     = {https://arxiv.org/abs/1308.3432}
}

@inproceedings{fang2024maskllm,
  title={Maskllm: Learnable semi-structured sparsity for large language models},
  author={Fang, Gongfan and Yin, Hongxu and Muralidharan, Saurav and Heinrich, Greg and Pool, Jeff and Kautz, Jan and Molchanov, Pavlo and Wang, Xinchao},
  booktitle={Advances in Neural Information Processing Systems (NeurIPS)},
  year={2024},
  url       = {https://arxiv.org/abs/2409.17481}

}

@inproceedings{frantar2025marlin,
author = {Frantar, Elias and Castro, Roberto L. and Chen, Jiale and Hoefler, Torsten and Alistarh, Dan},
title = {MARLIN: Mixed-Precision Auto-Regressive Parallel Inference on Large Language Models},
year = {2025},
url = {https://doi.org/10.1145/3710848.3710871},
booktitle = {Proceedings of the 30th ACM SIGPLAN Annual Symposium on Principles and Practice of Parallel Programming},
}

@article{dadgarnia2026gsq,
  title   = {{GSQ}: Highly-Accurate Low-Precision Scalar Quantization for {LLMs} via {Gumbel-Softmax} Sampling},
  author  = {Dadgarnia, Alireza and Tabesh, Soroush and Nikdan, Mahdi and Helcig, Michael and Kurtic, Eldar and Kleinegger, Maximilian and Alistarh, Dan},
  journal = {arXiv preprint arXiv:2604.18556},
  year    = {2026},
  url     = {https://arxiv.org/abs/2604.18556}
}

@inproceedings{rein2024gpqa,
title={{GPQA}: A Graduate-Level Google-Proof Q\&A Benchmark},
author={David Rein and Betty Li Hou and Asa Cooper Stickland and Jackson Petty and Richard Yuanzhe Pang and Julien Dirani and Julian Michael and Samuel R. Bowman},
booktitle={First Conference on Language Modeling},
year={2024},
url={https://openreview.net/forum?id=Ti67584b98}
}

@inproceedings{hendrycks2021math,
  title     = {Measuring Mathematical Problem Solving With the {MATH} Dataset},
  author={Hendrycks, Dan and Burns, Collin and Kadavath, Saurav and Arora, Akul and Basart, Steven and Tang, Eric and Song, Dawn and Steinhardt, Jacob},
  booktitle = {NeurIPS Datasets and Benchmarks Track},
  year      = {2021},
  url       = {https://arxiv.org/abs/2103.03874}
}

@misc{aime25,
      title={American Invitational Mathematics Examination (AIME) 2025}, 
      author={Zhang, Yifan and Math-AI, Team},
      year={2025},
}

@misc{cutlass_blackwell_docs,
  title         = {{CUTLASS} Functionality on {NVIDIA} {Blackwell} Architecture},
  author        = {{NVIDIA Corporation}},
  howpublished  = {NVIDIA developer documentation},
  year          = {2025},
  url           = {https://docs.nvidia.com/cutlass/latest/media/docs/cpp/blackwell_functionality.html}
}

@misc{nvidia2025cutlassgroupednvfp4,
  title        = {{Blackwell Grouped GEMM with Block-Scaled Narrow-Precision Inputs}},
  author       = {{NVIDIA Corporation}},
  howpublished = {NVIDIA CUTLASS example 75},
  year         = {2025},
  url          = {https://github.com/NVIDIA/cutlass/blob/main/examples/75_blackwell_grouped_gemm/75_blackwell_grouped_gemm_block_scaled.cu}
}

@misc{vllm2026kimik25recipe,
  title        = {{Kimi-K2.5} Deployment Recipe},
  author       = {{vLLM Project}},
  howpublished = {vLLM model recipe},
  year         = {2026},
  url          = {https://github.com/vllm-project/recipes/blob/main/models/moonshotai/Kimi-K2.5.yaml}
}

@misc{qutlass2025,
      title={QuTLASS: CUTLASS-Powered Quantized BLAS for Deep Learning},
      author={Castro, Roberto L. and Alistarh, Dan},
      year={2025},
      publisher = {GitHub},
      howpublished = {\url{https://github.com/IST-DASLab/qutlass}},
}

@inproceedings{shazeer2017moe,
  title     = {Outrageously Large Neural Networks: The Sparsely-Gated Mixture-of-Experts Layer},
  author    = {Shazeer, Noam and Mirhoseini, Azalia and Maziarz, Krzysztof and Davis, Andy and Le, Quoc and Hinton, Geoffrey and Dean, Jeff},
  booktitle = {International Conference on Learning Representations (ICLR)},
  year      = {2017},
  url       = {https://openreview.net/forum?id=B1ckMDqlg}
}

@inproceedings{lepikhin2021gshard,
  title     = {{GShard}: Scaling Giant Models with Conditional Computation and Automatic Sharding},
  author    = {Lepikhin, Dmitry and Lee, HyoukJoong and Xu, Yuanzhong and Chen, Dehao and Firat, Orhan and Huang, Yanping and Krikun, Maxim and Shazeer, Noam and Chen, Zhifeng},
  booktitle = {International Conference on Learning Representations (ICLR)},
  year      = {2021},
  url       = {https://openreview.net/forum?id=qrwe7XHTmYb}
}

@article{fedus2022switch,
  title   = {Switch Transformers: Scaling to Trillion Parameter Models with Simple and Efficient Sparsity},
  author  = {Fedus, William and Zoph, Barret and Shazeer, Noam},
  journal = {Journal of Machine Learning Research (JMLR)},
  year    = {2022},
  url     = {https://arxiv.org/abs/2101.03961}
}

@article{deepseekai2026v4,
  title   = {{DeepSeek-V4}: Towards Highly Efficient Million-Token Context Intelligence},
  author  = {{DeepSeek-AI}},
  journal = {arXiv preprint arXiv:2606.19348},
  year    = {2026},
  url     = {https://arxiv.org/abs/2606.19348},
}

@article{kimiteam2026k25,
  title   = {Kimi {K2.5}: Visual Agentic Intelligence},
  author  = {{Kimi Team}},
  journal = {arXiv preprint arXiv:2602.02276},
  year    = {2026},
  url     = {https://arxiv.org/abs/2602.02276},
}

@misc{moonshot2026k3,
  title        = {Kimi {K3} Tech Blog: Open Frontier Intelligence},
  author       = {{Moonshot AI}},
  year         = {2026},
  howpublished = {Moonshot AI blog},
  url          = {https://www.kimi.com/blog/kimi-k3}, 
}

@inproceedings{krajewski2024moescaling,
  title     = {Scaling Laws for Fine-Grained Mixture of Experts},
  author={Ludziejewski, Jan and Krajewski, Jakub and Adamczewski, Kamil and Pi{\'o}ro, Maciej and Krutul, Michal and Antoniak, Szymon and Ciebiera, Kamil and Kr{\'o}l, Krystian and Odrzyg{\'o}zdz, Tomasz and Sankowski, Piotr and others},
  booktitle = {International Conference on Machine Learning (ICML)},
  year      = {2024},
  url       = {https://arxiv.org/abs/2402.07871},
}

@inproceedings{abnar2025sparsity,
  title = 	 {Parameters vs {FLOP}s: Scaling Laws for Optimal Sparsity for Mixture-of-Experts Language Models},
  author =       {Abnar, Samira and Shah, Harshay and Busbridge, Dan and El-Nouby, Alaaeldin and Susskind, Joshua M. and Thilak, Vimal},
  booktitle = {International Conference on Machine Learning (ICML)},
  year = 	 {2025},
  url = 	 {https://proceedings.mlr.press/v267/abnar25a.html},  
}

@inproceedings{nakamura2026optimalsparsity,
  title     = {Optimal Sparsity of Mixture-of-Experts Language Models for Reasoning Tasks},
  author    = {Nakamura, Taishi and Ishikawa, Satoki and Kawamura, Masaki and Okamoto, Takumi and Nohara, Daisuke and Suzuki, Jun and Yokota, Rio},
  booktitle = {International Conference on Learning Representations (ICLR)},
  year      = {2026},
  url       = {https://arxiv.org/abs/2508.18672},
}

@inproceedings{frantar2023sparsegpt,
  title     = {{SparseGPT}: Massive Language Models Can Be Accurately Pruned in One-Shot},
  author    = {Frantar, Elias and Alistarh, Dan},
  booktitle = {International Conference on Machine Learning (ICML)},
  year      = {2023},
  url       = {https://proceedings.mlr.press/v202/frantar23a/frantar23a.pdf}
}

@inproceedings{sun2024wanda,
  title={A Simple and Effective Pruning Approach for Large Language Models},
  author={Mingjie Sun and Zhuang Liu and Anna Bair and J Zico Kolter},
  booktitle={International Conference on Learning Representations (ICLR)},
  year={2024},
  url={https://openreview.net/forum?id=PxoFut3dWW}
}

@inproceedings{lee2026elsa,
  title     = {The Unseen Frontier: Pushing the Limits of {LLM} Sparsity with Surrogate-Free {ADMM}},
  author    = {Lee, Kwanhee and Jang, Hyeondo and Lee, Dongyeop and Alistarh, Dan and Lee, Namhoon},
  booktitle = {International Conference on Learning Representations (ICLR)},
  year      = {2026},
  url       = {https://arxiv.org/abs/2510.01650}
}

@inproceedings{frantar2023gptq,
  title     = {{GPTQ}: Accurate Post-Training Quantization for Generative Pre-trained Transformers},
  author    = {Frantar, Elias and Ashkboos, Saleh and Hoefler, Torsten and Alistarh, Dan},
  booktitle = {International Conference on Learning Representations (ICLR)},
  year      = {2023},
  url       = {https://arxiv.org/abs/2210.17323}
}

@inproceedings{lin2024awq,
  title     = {{AWQ}: Activation-aware Weight Quantization for {LLM} Compression and Acceleration},
  author={Lin, Ji and Tang, Jiaming and Tang, Haotian and Yang, Shang and Chen, Wei-Ming and Wang, Wei-Chen and Xiao, Guangxuan and Dang, Xingyu and Gan, Chuang and Han, Song},
  booktitle = {Proceedings of Machine Learning and Systems (MLSys)},
  year      = {2024},
  url       = {https://arxiv.org/abs/2306.00978}
}

@inproceedings{tseng2024qtip,
  title     = {{QTIP}: Quantization with Trellises and Incoherence Processing},
  author    = {Tseng, Albert and Sun, Qingyao and Hou, David and De Sa, Christopher},
  booktitle = {Advances in Neural Information Processing Systems (NeurIPS)},
  year      = {2024},
  url       = {https://arxiv.org/abs/2406.11235},
  note      = {Spotlight.}
}

@inproceedings{harma2025interplay,
  title     = {Effective Interplay between Sparsity and Quantization: From Theory to Practice},
  author={Harma, Simla and Chakraborty, Ayan and Kostenok, Elizaveta and Mishin, Danila and Ha, Dongho and Falsafi, Babak and Jaggi, Martin and Liu, Ming and Oh, Yunho and Subramanian, Suvinay and others},
  booktitle = {International Conference on Learning Representations (ICLR)},
  year      = {2025},
  url       = {https://arxiv.org/abs/2405.20935},
}

@inproceedings{kim2026order,
  title     = {Prune-then-Quantize or Quantize-then-Prune? {U}nderstanding the Impact of Compression Order in Joint Model Compression},
  author    = {Kim, Minjun and Choi, Jaehyeon and Yang, Hyunwoo and Kim, Jongjin and Song, Jinho and Kang, U},
  booktitle = {International Conference on Learning Representations (ICLR)},
  year      = {2026},
  url       = {https://arxiv.org/abs/2603.18426}
}

@inproceedings{guo2024jsq,
  title     = {Compressing Large Language Models by Joint Sparsification and Quantization},
  author =       {Guo, Jinyang and Wu, Jianyu and Wang, Zining and Liu, Jiaheng and Yang, Ge and Ding, Yifu and Gong, Ruihao and Qin, Haotong and Liu, Xianglong},
  booktitle = {International Conference on Machine Learning (ICML)},
  year      = {2024},
  url       = {https://proceedings.mlr.press/v235/guo24g.html}
}

@inproceedings{guo2026obr,
  title     = {Optimal Brain Restoration for Joint Quantization and Sparsification of {LLM}s},
  author    = {Guo, Hang and Benini, Luca and Li, Yawei},
  booktitle = {International Conference on Learning Representations (ICLR)},
  year      = {2026},
  url       = {https://openreview.net/forum?id=VQIvBpL5ag}
}

@inproceedings{lasby2025reap,
 author = {Lasby, Mike and Lazarevich, Ivan and Sinnadurai, Nish and Lie, Sean and Ioannou, Yani and Thangarasa, Vithursan},
 booktitle = {International Conference on Learning Representations (ICLR)},
 title = {REAP the Experts: Why Pruning Prevails for One-Shot MoE compression},
 url = {https://proceedings.iclr.cc/paper_files/paper/2026/file/ed93b2b5722acc2341d421b8916404a1-Paper-Conference.pdf},
 year = {2026}
}

@inproceedings{huang2025mcmoe,
  title     = {Mixture Compressor for Mixture-of-Experts {LLMs} Gains More},
  author={Huang, Wei and Liao, Yue and Liu, Jianhui and He, Ruifei and Tan, Haoru and Zhang, Shiming and Li, Hongsheng and Liu, Si and Qi, Xiaojuan},
  booktitle = {International Conference on Learning Representations (ICLR)},
  year      = {2025},
  url       = {https://arxiv.org/abs/2410.06270}
}

@article{fu2025eaquant,
  title   = {{EAQuant}: Enhancing Post-Training Quantization for {MoE} Models via Expert-Aware Optimization},
  author={Fu, Zhongqian and Zhao, Tianyi and Ding, Ning and Yu, Xianzhi and Li, Xiaosong and Tang, Yehui and Wang, Yunhe},
  journal = {arXiv preprint arXiv:2506.13329},
  year    = {2025},
  url     = {https://arxiv.org/abs/2506.13329}
}

@article{mishra2021nm,
  title   = {Accelerating Sparse Deep Neural Networks},
  author={Mishra, Asit and Latorre, Jorge Albericio and Pool, Jeff and Stosic, Darko and Stosic, Dusan and Venkatesh, Ganesh and Yu, Chong and Micikevicius, Paulius},
  journal = {arXiv preprint arXiv:2104.08378},
  year    = {2021},
  url     = {https://arxiv.org/abs/2104.08378},
}

@article{rouhani2023mx,
  title   = {Microscaling Data Formats for Deep Learning},
  author={Rouhani, Bita Darvish and Zhao, Ritchie and More, Ankit and Hall, Mathew and Khodamoradi, Alireza and Deng, Summer and Choudhary, Dhruv and Cornea, Marius and Dellinger, Eric and Denolf, Kristof and others},
  journal = {arXiv preprint arXiv:2310.10537},
  year    = {2023},
  url     = {https://arxiv.org/abs/2310.10537},
  note    = {OCP Microscaling (MX) format specification.}
}

@article{nvidia2025nvfp4,
  title   = {Pretraining Large Language Models with {NVFP4}},
  author  = {{NVIDIA}},
  journal = {arXiv preprint arXiv:2509.25149},
  year    = {2025},
  url     = {https://arxiv.org/abs/2509.25149}
}

@inproceedings{egiazarian2026bridging,
  title={Bridging the gap between promise and performance for microscaling fp4 quantization},
  author={Egiazarian, Vage and Castro, Roberto and Kuznedelev, Denis and Panferov, Andrei and Kurtic, Eldar and Pandit, Shubhra and Marques, Alexandre and Kurtz, Mark and Ashkboos, Saleh and Hoefler, Torsten and others},
  booktitle={International Conference on Learning Representations (ICLR)},
  volume={2026},
  year={2026}
}

@misc{zai2026glm52,
  title        = {{GLM-5.2}},
  author       = {{Z.ai}},
  year         = {2026},
  howpublished = {Hugging Face model card},
  url          = {https://huggingface.co/zai-org/GLM-5.2},
}

@misc{lgai2026kexaone2,
  title        = {{K-EXAONE 2.0}},
  author       = {{LG AI Research}},
  year         = {2026},
  howpublished = {Hugging Face model card},
  url          = {https://huggingface.co/LGAI-EXAONE/K-EXAONE-2.0-750B-A37B},
}

@misc{qwen2026qwen38,
    title = {{Qwen3.8-Max}: A New Bar for Coding and Cowork},
    url = {https://qwen.ai/blog?id=qwen3.8},
    author = {{Qwen Team}},
    month = {August},
    year = {2026}
}

@misc{qwen2026qwen35,
  title        = {{Qwen3.5-397B-A17B}},
  author       = {{Qwen Team}},
  year         = {2026},
  howpublished = {Hugging Face model card},
  url          = {https://huggingface.co/Qwen/Qwen3.5-397B-A17B},
}

@misc{nvidia2026kimik25nvfp4,
  title        = {{Kimi-K2.5-NVFP4}},
  author       = {{NVIDIA Corporation}},
  year         = {2026},
  howpublished = {Hugging Face model card},
  url          = {https://huggingface.co/nvidia/Kimi-K2.5-NVFP4},
}

@misc{nvidia2026qwen35nvfp4,
  title        = {{Qwen3.5-397B-A17B-NVFP4}},
  author       = {{NVIDIA Corporation}},
  year         = {2026},
  howpublished = {Hugging Face model card},
  url          = {https://huggingface.co/nvidia/Qwen3.5-397B-A17B-NVFP4},
}

@misc{nvidia2025qwen330bnvfp4,
  title        = {{Qwen3-30B-A3B-NVFP4}},
  author       = {{NVIDIA Corporation}},
  year         = {2025},
  howpublished = {Hugging Face model card},
  url          = {https://huggingface.co/nvidia/Qwen3-30B-A3B-NVFP4},
}

@misc{ocp2023mx,
  title        = {{OCP} Microscaling Formats ({MX}) Specification, Version 1.0},
  author       = {{Open Compute Project Foundation (MX Alliance)}},
  year         = {2023},
  howpublished = {Open Compute Project specification},
  url          = {https://www.opencompute.org/documents/ocp-microscaling-formats-mx-v1-0-spec-final-pdf}
}

@misc{humming2026,
  title        = {{Humming}: A {JIT}-Compiled {GEMM} Kernel Library for Quantized Inference},
  author       = {{inclusionAI}},
  year         = {2026},
  publisher    = {GitHub},
  howpublished = {\url{https://github.com/inclusionAI/humming}},
}

@inproceedings{chen2023lion,
  title     = {Symbolic Discovery of Optimization Algorithms},
  author={Chen, Xiangning and Liang, Chen and Huang, Da and Real, Esteban and Wang, Kaiyuan and Pham, Hieu and Dong, Xuanyi and Luong, Thang and Hsieh, Cho-Jui and Lu, Yifeng and others},
  booktitle = {Advances in Neural Information Processing Systems (NeurIPS)},
  year      = {2023},
  url       = {https://arxiv.org/abs/2302.06675}
}

@misc{neuralmagic2024llmcompressioncalibration,
  title        = {{LLM Compression Calibration}},
  author       = {{Neural Magic}},
  year         = {2024},
  month        = jun,
  howpublished = {Hugging Face dataset},
  url          = {https://huggingface.co/datasets/neuralmagic/LLM_compression_calibration}
}

@inproceedings{guha2025openthoughtsdatarecipesreasoning,
  title={OpenThoughts: Data Recipes for Reasoning Models}, 
  author={Etash Guha and Ryan Marten and Sedrick Keh and Negin Raoof and Georgios Smyrnis and Hritik Bansal and Marianna Nezhurina and Jean Mercat and Trung Vu and Zayne Sprague and Ashima Suvarna and Benjamin Feuer and Liangyu Chen and Zaid Khan and Eric Frankel and Sachin Grover and Caroline Choi and Niklas Muennighoff and Shiye Su and Wanjia Zhao and John Yang and Shreyas Pimpalgaonkar and Kartik Sharma and Charlie Cheng-Jie Ji and Yichuan Deng and Sarah Pratt and Vivek Ramanujan and Jon Saad-Falcon and Jeffrey Li and Achal Dave and Alon Albalak and Kushal Arora and Blake Wulfe and Chinmay Hegde and Greg Durrett and Sewoong Oh and Mohit Bansal and Saadia Gabriel and Aditya Grover and Kai-Wei Chang and Vaishaal Shankar and Aaron Gokaslan and Mike A. Merrill and Tatsunori Hashimoto and Yejin Choi and Jenia Jitsev and Reinhard Heckel and Maheswaran Sathiamoorthy and Alexandros G. Dimakis and Ludwig Schmidt},
  booktitle={International Conference on Learning Representations (ICLR)},
  year={2026},
  url={https://arxiv.org/abs/2506.04178}, 
}

@misc{lozhkov2024fineweb-edu,
    author       = { Lozhkov, Anton and Ben Allal, Loubna and von Werra, Leandro and Wolf, Thomas },  
    title        = { FineWeb-Edu: the Finest Collection of Educational Content }, 
    year         = 2024,  
    url          = { https://huggingface.co/datasets/HuggingFaceFW/fineweb-edu },  
    doi          = { 10.57967/hf/2497 },
    publisher    = { Hugging Face }
}

@misc{nvidia2021a100datasheet,
  title        = {{NVIDIA} {A100} Tensor Core {GPU}: Unprecedented Acceleration at Every Scale},
  author       = {{NVIDIA Corporation}},
  howpublished = {NVIDIA datasheet},
  month        = jun,
  year         = {2021},
  url          = {https://www.nvidia.com/content/dam/en-zz/Solutions/Data-Center/a100/pdf/nvidia-a100-datasheet-us-nvidia-1758950-r4-web.pdf}
}

@misc{nvidia2022h100datasheet,
  title        = {{NVIDIA} {H100} Tensor Core {GPU} Datasheet},
  author       = {{NVIDIA Corporation}},
  howpublished = {NVIDIA datasheet},
  year         = {2022},
  url          = {https://resources.nvidia.com/en-us-gpu-resources/h100-datasheet-24306}
}

@misc{nvidia2024dgxb200datasheet,
  title        = {{NVIDIA} {DGX} {B200} Datasheet},
  author       = {{NVIDIA Corporation}},
  howpublished = {NVIDIA datasheet},
  year         = {2024},
  url          = {https://resources.nvidia.com/en-us-dgx-systems/dgx-b200-datasheet}
}

@inproceedings{kwon2023vllm,
  title     = {Efficient Memory Management for Large Language Model Serving with {PagedAttention}},
  author={Kwon, Woosuk and Li, Zhuohan and Zhuang, Siyuan and Sheng, Ying and Zheng, Lianmin and Yu, Cody Hao and Gonzalez, Joseph and Zhang, Hao and Stoica, Ion},
  booktitle = {Proceedings of the 29th Symposium on Operating Systems Principles (SOSP)},
  year      = {2023},
  url       = {https://arxiv.org/abs/2309.06180}
}

@inproceedings{ye2025flashinfer,
  title     = {{FlashInfer}: Efficient and Customizable Attention Engine for {LLM} Inference Serving},
  author={Ye, Zihao and Chen, Lequn and Lai, Ruihang and Lin, Wuwei and Zhang, Yineng and Wang, Stephanie and Chen, Tianqi and Kasikci, Baris and Grover, Vinod and Krishnamurthy, Arvind and others},
  booktitle = {Proceedings of Machine Learning and Systems (MLSys)},
  year      = {2025},
  url       = {https://arxiv.org/abs/2501.01005}
}

@misc{hejazi2024groupedgemm,
  title        = {Introducing Grouped {GEMM} {API}s in {cuBLAS} and More Performance Updates},
  author       = {Hejazi, Babak},
  howpublished = {NVIDIA Technical Blog},
  month        = jun,
  year         = {2024},
  note         = {Published June 12, 2024; accessed 2026-09-22},
  url          = {https://developer.nvidia.com/blog/introducing-grouped-gemm-apis-in-cublas-and-more-performance-updates/}
}

@inproceedings{lucas2025reasoningpruning,
  title   = {Reasoning Models Can Be Accurately Pruned via Chain-of-Thought Reconstruction},
  author  = {Lucas, Ryan and Behdin, Kayhan and Wang, Zhipeng and Song, Qingquan and Tang, Shao and Mazumder, Rahul},
  booktitle = {International Conference on Learning Representations (ICLR)},
  year    = {2026},
  doi     = {10.48550/arXiv.2509.12464},
  url     = {https://arxiv.org/abs/2509.12464}
}
